\documentclass[runningheads]{llncs}

\usepackage[T1]{fontenc}
\usepackage{graphicx}
\usepackage{xcolor}
\usepackage{xspace}
\usepackage{color}
\usepackage{amsmath}
\usepackage{hyperref}

\begin{document}


\newenvironment{compactitem}%
{\begin{itemize}%
		\setlength{\itemsep}{0pt}%
		\setlength{\parskip}{0pt}}%
{\end{itemize}}

\newcommand{\myi}{(\emph{i})\xspace}
\newcommand{\myii}{(\emph{ii})\xspace}
\newcommand{\myiii}{(\emph{iii})\xspace}
\newcommand{\myiv}{(\emph{iv})\xspace}
\newcommand{\myv}{(\emph{v})\xspace}
\newcommand{\myvi}{(\emph{vi})\xspace}
\newcommand{\myvii}{(\emph{vii})\xspace}
\newcommand{\myviii}{(\emph{viii})\xspace}

\newcommand{\A}{\mathcal{A}} \newcommand{\B}{\mathcal{B}}
\newcommand{\C}{\mathcal{C}} \newcommand{\D}{\mathcal{D}}
\newcommand{\E}{\mathcal{E}} \newcommand{\F}{\mathcal{F}}
\newcommand{\G}{\mathcal{G}} \renewcommand{\H}{\mathcal{H}}
\newcommand{\I}{\mathcal{I}} \newcommand{\J}{\mathcal{J}}
\newcommand{\K}{\mathcal{K}} \renewcommand{\L}{\mathcal{L}}
\newcommand{\M}{\mathcal{M}} \newcommand{\N}{\mathcal{N}}
\renewcommand{\O}{\mathcal{O}} \renewcommand{\P}{\mathcal{P}}
\newcommand{\Q}{\mathcal{Q}} \newcommand{\R}{\mathcal{R}}
\renewcommand{\S}{\mathcal{S}} \newcommand{\T}{\mathcal{T}}
\newcommand{\U}{\mathcal{U}} \newcommand{\V}{\mathcal{V}}
\newcommand{\W}{\mathcal{W}} \newcommand{\X}{\mathcal{X}}
\newcommand{\Y}{\mathcal{Y}} \newcommand{\Z}{\mathcal{Z}}
\newcommand{\DB}{\D\B}


\newcommand{\setone}[2][1]{\set{#1\cld #2}}
\newcommand{\eset}{\emptyset}
\newcommand{\ol}[1]{\overline{#1}}                
\newcommand{\ul}[1]{\underline{#1}}               
\newcommand{\uls}[1]{\underline{\raisebox{0pt}[0pt][0.45ex]{}#1}}
\newcommand{\wrt}{w.r.t.\xspace}
\newcommand{\ra}{\rightarrow}
\newcommand{\Ra}{\Rightarrow}
\newcommand{\la}{\leftarrow}
\newcommand{\La}{\Leftarrow}
\newcommand{\lra}{\leftrightarrow}
\newcommand{\Lra}{\Leftrightarrow}
\newcommand{\lora}{\longrightarrow}
\newcommand{\Lora}{\Longrightarrow}
\newcommand{\lola}{\longleftarrow}
\newcommand{\Lola}{\Longleftarrow}
\newcommand{\lolra}{\longleftrightarrow}
\newcommand{\Lolra}{\Longleftrightarrow}
\newcommand{\ua}{\uparrow}
\newcommand{\Ua}{\Uparrow}
\newcommand{\da}{\downarrow}
\newcommand{\Da}{\Downarrow}
\newcommand{\uda}{\updownarrow}
\newcommand{\Uda}{\Updownarrow}


\newcommand{\incl}{\subseteq}
\newcommand{\imp}{\rightarrow}
\newcommand{\dleq}{\dot{\leq}}                   


\newcommand{\per}{\mbox{\bf .}}                  

\newcommand{\cld}{,\ldots,}                      
\newcommand{\ld}[1]{#1 \ldots #1}                 
\newcommand{\cd}[1]{#1 \cdots #1}                 
\newcommand{\lds}[1]{\, #1 \; \ldots \; #1 \,}    
\newcommand{\cds}[1]{\, #1 \; \cdots \; #1 \,}    

\newcommand{\dd}[2]{#1_1,\ldots,#1_{#2}}             
\newcommand{\ddd}[3]{#1_{#2_1},\ldots,#1_{#2_{#3}}}  
\newcommand{\dddd}[3]{#1_{11}\cld #1_{1#3_{1}}\cld #1_{#21}\cld #1_{#2#3_{#2}}}

\newcommand{\ldop}[3]{#1_1 \ld{#3} #1_{#2}}   
\newcommand{\cdop}[3]{#1_1 \cd{#3} #1_{#2}}   
\newcommand{\ldsop}[3]{#1_1 \lds{#3} #1_{#2}} 
\newcommand{\cdsop}[3]{#1_1 \cds{#3} #1_{#2}} 


\newcommand{\quotes}[1]{{\lq\lq #1\rq\rq}}
\newcommand{\set}[1]{\{#1\}}                      
\newcommand{\Set}[1]{\left\{#1\right\}}
\newcommand{\bigmid}{\Big|}
\newcommand{\card}[1]{|{#1}|}                     
\newcommand{\Card}[1]{\left| #1\right|}
\newcommand{\cards}[1]{\sharp #1}
\newcommand{\sub}[1]{[#1]}
\newcommand{\tup}[1]{\langle #1\rangle}            
\newcommand{\Tup}[1]{\left\langle #1\right\rangle}
\newcommand{\inc}[2]{#1\colon #2}

\newcommand{\quonto}{\textsc{QuOnto}\xspace}
\newcommand{\dllite}{\textit{DL-Lite}\xspace}
\newcommand{\dlliter}{\textit{DL-Lite}_{R}\xspace}
\newcommand{\dllitef}{\textit{DL-Lite}_{F}\xspace}
\newcommand{\dllitefr}{\textit{DL-Lite}_{F\R}\xspace}
\newcommand{\dllitea}{\textit{DL-Lite}_{A}\xspace}
\newcommand{\dlliteaid}{\textit{DL-Lite}_{A,id}\xspace}
\newcommand{\dllitefrs}{{{\textit{DL-Lite}}_{\mathit{FRS}}}\xspace}
\newcommand{\dllitefrcmin}{{{\textit{DL-Lite}}_{\mathiti{FR}}}\xspace}
\newcommand{\dlrliter}{{\textit{DLR-Lite}_{\R}}\xspace}
\newcommand{\dllitefs}{{{\textit{DL-Lite}}_{\mathit{FS}}}\xspace}
\newcommand{\dllitefa}{\textit{DL-Lite}_{\mathit{FA}}\xspace}
\newcommand{\dlrliteand}{\textit{DLR-Lite$_{\mathcal{A},\AND}$}\xspace}
\newcommand{\dlliteaidden}{\textit{DL-Lite}_{A,id,den}\xspace}
\newcommand{\dlliteden}{\textit{DL-Lite}_{A,id,den}\xspace}
\newcommand{\LOGSPACE}{\textsc{LogSpace}\xspace}
\newcommand{\NLOGSPACE}{\textsc{NLogSpace}\xspace}
\newcommand{\PTIME}{\textsc{PTime}\xspace}
\newcommand{\EXPTIME}{\textsc{ExpTime}\xspace}
\newcommand{\NEXPTIME}{\textsc{NExpTime}\xspace}
\newcommand{\NP}{\textrm{NP}\xspace}
\newcommand{\NPcomp}{\textrm{NP-Complete}\xspace}
\newcommand{\NPhard}{\textrm{NP-Hard}\xspace}
\newcommand{\coNP}{\textrm{coNP}\xspace}
\newcommand{\ACz}{\textsc{AC}\ensuremath{^0}\xspace}

\newcommand{\ALCQIb}{\mathcal{ALCQI}b}
\newcommand{\AL}{\mathcal{AL}}
\newcommand{\ALE}{\mathcal{ALE}}
\newcommand{\ALU}{\mathcal{ALU}}
\newcommand{\FLE}{\mathcal{FLE}}
\newcommand{\EL}{\mathcal{EL}}
\newcommand{\SHIQ}{\mathcal{SHIQ}}
\newcommand{\ALCQIO}{\mathcal{ALCQIO}}


\newcommand{\AND}{\sqcap}
\newcommand{\OR}{\sqcup}
\newcommand{\NOT}{\neg}
\newcommand{\ALL}[2]{\forall #1 \per #2}
\newcommand{\SOME}[2]{\exists #1 \per #2}
\newcommand{\SOMET}[1]{\exists #1}
\newcommand{\ALLRC}{\ALL{R}{C}}
\newcommand{\SOMERC}{\SOME{R}{C}}
\newcommand{\SOMERT}{\SOMET{R}}
\newcommand{\ATLEAST}[2]{(\geq #1 \, #2)}
\newcommand{\ATMOST}[2]{(\leq #1 \, #2)}
\newcommand{\EXACTLY}[2]{(= #1 \, #2)}
\newcommand{\INV}[1]{#1^{-}}

\newcommand{\atom}[3][]{\gamma_{#1}(#2,#3)}
\newcommand{\ISA}{\sqsubseteq}
\newcommand{\EQU}{\equiv}
\newcommand{\IMP}{\mathop{\Ra}}
\newcommand{\DIFF}{\setminus}
\newcommand{\COMP}[2]{\$#1\mathop{:}#2}
\newcommand{\QATLEAST}[3]{(\geq #1\,#2 \per\, #3)}
\newcommand{\QATMOST}[3]{(\leq #1\,#2 \per\, #3)}
\newcommand{\ID}{\mathit{id}}
\newcommand{\SOMER}[2]{\exists[\$#1]#2}
\newcommand{\ATMOSTR}[3]{(\leq #1\, [\$#2]#3)}
\newcommand{\PROJ}[3]{#1|_{#2,#3}}
\newcommand{\head}{\mathit{head}}
\newcommand{\body}{\mathit{body}}
\newcommand{\rel}[1]{\mathit{#1}}
\newcommand{\bodyl}[1][j]{\body_{#1}(\vett{a},\vett{b}_{#1},\vett{c}_{#1})}
\newcommand{\bodyr}[1][j]{\lnot\exists\vett{z}_{#1}\per
                          \body'_{#1}(\vett{a},\vett{z}_{#1},\vett{c}_{#1})}
\newcommand{\arity}[1]{\mathit{arity}(#1)}
\newcommand{\vett}[1]{\vec{\mathbf{#1}}}
\newcommand{\intsys}{\langle\G,\S,\M\rangle}
\newcommand{\true}{\textit{true}}
\newcommand{\false}{\textit{false}}
\newcommand{\vv}{\vec{v}}
\newcommand{\SemIcar}[1]{\mathit{Sem}_{\mathit{ICAR}}(#1)}
\newcommand{\ModIcar}[1]{\mathit{Mod}_{\mathit{ICAR}}(#1)}
\newcommand{\chase}{\mathit{chase}}
\newcommand{\unfold}{\mathit{unfold}}
\newcommand{\mapp}{\rho}
\newcommand{\CSP}{{\sc CSP}}
\newcommand{\UCQ}{{\sc UCQ}\xspace}
\newcommand{\UCQs}{{\sc UCQ}s\xspace}
\newcommand{\ie}{\emph{i.e.,}\xspace}
\newcommand{\eg}{\emph{e.g.,}\xspace}
\newcommand{\conj}{\textit{conj}}
\newcommand{\gr}{\textit{ngr}}
\newcommand{\trasf}{\tau}
\newcommand{\ass}{\leftarrow}
\newcommand{\idrewr}{\mathsf{IDrewrite}}
\newcommand{\funct}[1]{(\mathsf{funct}\; #1)}
\newcommand{\functir}[2]{(\mathsf{funct}\; #1{:}#2)}
\newcommand{\FUNCT}[1]{(\mathsf{funct}\; #1)}
\newcommand{\KEYR}[2]{(\mathsf{key}\; #1{:}\,#2)}
\newcommand{\SOMEIRQ}[3]{\exists #1{:}#2\per #3}
\newcommand{\range}[2]{(\mathsf{range}\; #1~#2)}
\newcommand{\se}{\noindent \leftarrow}
\newcommand{\Ans}{\mathit{ans}}
\newcommand{\tab}{\mathit{tab}}
\newcommand{\aux}{\mathit{aux}}
\newcommand{\SOMEIR}[2]{\exists #1{:}#2}
\newcommand{\IR}[2]{#1{:}#2}
\newcommand{\DOMAIN}[1]{\delta(#1)}
\newcommand{\RANGE}[1]{\rho(#1)}
\newcommand{\DOMAINSEL}[2]{\delta_{#1}(#2)}
\newcommand{\topc}{\top_C}
\newcommand{\topd}{\top_D}
\newcommand{\Map}[3]{ (#1,#2,#3)}
\newcommand{\limp}{\ra}

\newcommand{\inter}[1][I]{(\dom[#1],\Int[#1]{\cdot})}   
\newcommand{\dom}[1][I]{\Delta^{#1}}  
\newcommand{\domO}[1][~I]{\Delta_O^{#1}}  
\newcommand{\domV}[1][~I]{\Delta_V^{#1}}  
\newcommand{\fint}[1][I]{\Int[#1]{\cdot}}
\newcommand{\Int}[2][I]{#2^{#1}}      
\newcommand{\INT}[2][I]{(#2)^{#1}}    
\newcommand{\IntDB}[2][\DB]{#2^{#1}}      
\newcommand{\intsysdl}{\langle\G,\C,\S,\M\rangle}
\newcommand{\IS}{\Pi}
\newcommand{\perfectref}{\textsf{PerfectRef}}
\newcommand{\assign}[1][]{\mu^{#1}}
\newcommand{\assigni}[3]{{#1}[{#2}/{#3}]}
\newcommand{\clni}[1][i]{\mathit{cln}_{#1}(\T)}
\newcommand{\NSFR}{NSFR}
\newcommand{\NI}{NI\xspace}
\newcommand{\PI}{PI\xspace}
\newcommand{\unb}{\texttt{\_}\,}
\newcommand{\crule}[1]{\textbf{cr{#1}}}
\newcommand{\dbi}[2]{\mathit{db}(#1,#2)}
\newcommand{\basic}{\textsf{KBBasicRef}}
\newcommand{\unifiedKB}{\textsf{KBUnifiedRef\xspace}}
\newcommand{\violni}{\textsf{ViolateNI}}
\newcommand{\violf}{\textsf{ViolateFunct}}
\newcommand{\assignO}{{\mu_{\bf 0}}}
\newcommand{\assignOi}[2][\assignO]{{#2}[{#1}]}
\newcommand{\Identity}{\mathit{Id}}
\newcommand{\dlliteam}{\textit{DL-Lite}_{A,M}\xspace}
\newcommand{\model}[1][M]{\inter[{#1}]}
\newcommand{\relation}[1]{\mathsf{{#1}}\xspace}
\newcommand{\rewrite}{\mathsf{RewDB}\xspace}
\newcommand{\modelsat}{\mathsf{ModelSat}\xspace}
\newcommand{\modelreach}{\mathsf{ModelReach}\xspace}
\newcommand{\clean}{\mathsf{Clean}\xspace}
\newcommand{\prdb}{\mathit{PR^{DB}}}
\newcommand{\conv}{\textsf{Conv}}
\newcommand{\GammaO}{\Gamma_O}
\newcommand{\GammaV}{\Gamma_V}
\newcommand{\modO}{I_{\bf 0}}
\newcommand{\VO}{\V_O}
\newcommand{\VV}{\V_V}
\newcommand{\bd}{\D\B}
\newcommand{\modelAss}[1][\assign]{\models_{#1}}
\newcommand{\subst}{\sigma}
\newcommand{\vx}{\vec{x}}
\newcommand{\vy}{\vec{y}}
\newcommand{\vd}{\vec{d}}
\newcommand{\vfx}{\vec{f(x)}}
\newcommand{\vfy}{\vec{f(y)}}
\newcommand{\vfd}{\vec{f(d)}}
\newcommand{\splitted}{\textsf{Split}}
\newcommand{\sld}{\textsf{SLD-Derive}}
\newcommand{\violates}{\textsf{Violates}}
\newcommand{\size}{\textsf{size}}
\newcommand{\TEST}[1]{#1?}
\newcommand{\rma}[3]{#1(#2,#3)}
\newcommand{\length}{\mathit{length}}

\newcommand{\ans}[2]{\mathit{#1^{#2}}}
\newcommand{\ansx}[2]{\mathit{#1^{#2}}}
\newcommand{\ansdb}[2]{\mathit{ans(#1,#2)}}
\newcommand{\mgu}{\textit{mgu}\xspace}
\newcommand{\ontmap}{\langle \T, \M, \DB \rangle}
\newcommand{\ontmapi}{\langle \T', \M', \DB' \rangle}
\newcommand{\OM}{{\O_{m}}}
\newcommand{\db}[1][\A]{{\mathit{DB}({#1})}}
\newcommand{\cln}[1][\T]{\mathit{cln}(#1)}
\newcommand{\PerfectRef}{\textsf{PerfectRef}\xspace}
\newcommand{\pr}{\mathit{PR}}
\newcommand{\Sat}{\textsf{Sat}\xspace}
\newcommand{\Answer}{\textsf{Answer}}
\newcommand{\alltuples}{\textit{AllTup}}
\newcommand{\vt}{\vec{t}}
\newcommand{\vo}{\vec{o}}
\newcommand{\vxi}{\vec{x'}}
\newcommand{\directDIS}{\textsf{DirectDIS}}
\newcommand{\ComputeVSets}{\textsf{ComputeVSets}\xspace}
\newcommand{\SatisfiableIDC}{\textsf{SatisfiableIDC}\xspace}
\newcommand{\Fillers}{\textsf{Fillers}\xspace}
\newcommand{\Clash}{\emph{Clash}\xspace}
\newcommand{\upd}{\circ}
\newcommand{\updateM}[3][\T]{{#2}\text{\boldmath $\circ$}_{#1}{#3}}
\newcommand{\update}[3][\T]{{#2}\circ_{#1}{#3}}
\newcommand{\upmod}[3][\T]{U^{#1}({#2},{#3})}
\newcommand{\simdiff}{\ominus}
\newcommand{\resup}[3]{\mathit{ComputeUpdate}({#1},{#2},{#3})}
\newcommand{\resupshort}[0]{\mathit{ComputeUpdate}}
\newcommand{\sat}[3][\T]{\mathit{SatModel}({#1},{#2},{#3})}
\newcommand{\ALC}{\ensuremath{\mathcal{ALC}}}


\newcommand{\QuOnto}{\textrm{QuOnto}\xspace}
\newcommand{\styx}{\textrm{StyX}\xspace}
\newcommand{\carin}{\textrm{CARIN}\xspace}
\newcommand{\picsel}{\textrm{PICSEL}\xspace}
\newcommand{\MASTROI}{\textsc{Mastro-i}\xspace}


\newcommand{\HB}{\textit{HB}\xspace}

\newcommand{\mcsar}{MCS_{\T}^{AR}\xspace}
\newcommand{\mcsiar}{MCS_{\T}^{IAR}\xspace}
\newcommand{\mcscar}{MCS_{\T}^{CAR}\xspace}
\newcommand{\mcsicar}{MCS_{\T}^{ICAR}\xspace}
\newcommand{\mcsx}{MCS_{\T}^{X}\xspace}

\newcommand{\carrepabox}{\textit{CA-rep}\xspace}
\newcommand{\oar}{OA\xspace}
\newcommand{\cox}{CX\xspace}
\newcommand{\crew}{CR\xspace}
\newcommand{\mof}{mf\xspace}
\newcommand{\fb}{fb\xspace}
\newcommand{\rtm}{RTM\xspace}
\newcommand{\ch}{CH\xspace}
\newcommand{\ath}{ATH\xspace}


\newcommand{\dllitecore}{\textit{DL-Lite}_{\mathit{core}}}
\newcommand{\elabel}{\ell}
\newcommand{\colorability}{\mathit{col}}

\newcommand{\carepair}{\mathit{CA}\rm{-repair}\xspace}

\newcommand{\modelsar}{\models_{\ar}}
\newcommand{\modelsiar}{\models_{\iar}}
\newcommand{\modelscar}{\models_{\car}}
\newcommand{\modelsicar}{\models_{\icar}}

\newcommand{\ia}{\alpha}
\newcommand{\Del}{\D}
\newcommand{\computeiarrepair}{\textit{Compute-IAR-Repair}}
\newcommand{\computeicarrepair}{\textit{Compute-ICAR-Repair}}

\newcommand{\CL}{\mathit{Cl}}
\newcommand{\CR}{\mathit{Cr}}
\newcommand{\QL}{\mathit{Ql}}
\newcommand{\QR}{\mathit{Qr}}
\newcommand{\VL}{\mathit{Vl}}
\newcommand{\VR}{\mathit{Vr}}


\newcommand{\obdm}{\textsf{ODIS}\xspace}
\newcommand{\ois}{\textsf{OIS}\xspace}
\newcommand{\sois}{\textsf{SOIS}\xspace}
\newcommand{\odis}{\textsf{ODIS}\xspace}
\newcommand{\sot}{SOT\xspace}

\newcommand{\Tinc}{\T_{inc}}
\newcommand{\Ttype}{\T_{type}}
\newcommand{\Tdisj}{\T_{disj}}
\newcommand{\Tfunct}{\T_{funct}}
\newcommand{\Tid}{\T_{id}}
\newcommand{\Tden}{\T_{den}}

\newcommand{\Alphabet}{\Gamma}

\newcommand{\arrep}{\textit{AR-Set}\xspace}
\newcommand{\armod}{\textit{Mod}_{AR}\xspace}
\newcommand{\iarrep}{\textit{IAR-Set}\xspace}
\newcommand{\iarmod}{\textit{Mod}_{IAR}\xspace}
\newcommand{\carrep}{\textit{CAR-Set}\xspace}
\newcommand{\carmod}{\textit{Mod}_{CAR}\xspace}
\newcommand{\icarrep}{\textit{ICA-Rep}\xspace}
\newcommand{\icarmod}{\textit{Mod}_{ICAR}\xspace}
\newcommand{\armodels}{\models_\textit{AR}\xspace}
\newcommand{\carmodels}{\models_\textit{CAR}\xspace}
\newcommand{\iarmodels}{\models_\textit{IAR}\xspace}
\newcommand{\icarmodels}{\models_\textit{ICAR}\xspace}
\newcommand{\car}{\textit{CA}\xspace}
\newcommand{\ar}{\textit{A}\xspace}
\newcommand{\iar}{\textit{IA}\xspace}
\newcommand{\icar}{\textit{ICA}\xspace}
\newcommand{\CAR}{\textit{CAR}\xspace}
\newcommand{\AR}{\textit{AR}\xspace}
\newcommand{\IAR}{\textit{IAR}\xspace}
\newcommand{\ICAR}{\textit{ICAR}\xspace}

\newcommand{\ComputeInsertion}{\textsf{ComputeInsertion}}
\newcommand{\ComputeDeletion}{\textsf{ComputeDeletion}}
\newcommand{\IncSets}{\textsf{InconsistentSets}}
\newcommand{\delete}{\textsf{Delete}}

\newcommand{\NaiveComputeAnswersar}{\textsf{NaiveComputeAnswers}_{AR}}
\newcommand{\NaiveComputeAnswerscar}{\textsf{NaiveComputeAnswers}_{CAR}}
\newcommand{\NaiveComputeAnswersiar}{\textsf{NaiveComputeAnswers}_{IAR}}
\newcommand{\NaiveComputeAnswersicar}{\textsf{NaiveComputeAnswers}_{ICAR}}
\newcommand{\facts}{\textsf{facts}}
\newcommand{\naiveupdateicar}{\textsf{NaiveComputeUpdate}_{ICAR}}
\newcommand{\computearrep}{\textsf{ComputeARSet}}
\newcommand{\computecarrep}{\textsf{ComputeCARSet}}
\newcommand{\computeiarrep}{\textsf{ComputeIARSet}}
\newcommand{\computeicarrep}{\textsf{ComputeICARSet}}
\newcommand{\computeclc}{\textsf{ComputeCLC}}
\newcommand{\images}{\textsf{images}}
\newcommand{\answericar}{\textsf{ComputeAnswer}_{ICAR}}
\newcommand{\updateicar}{\textsf{ComputeUpdate}_{ICAR}}
\newcommand{\answerDA}{\textsf{Answer}_{DA}}
\newcommand{\answer}{\textsf{Answer}}
\newcommand{\clash}[1]{#1\text{-}clash}
\newcommand{\ca}{\textit{CA}\xspace}
\newcommand{\insop}{\oplus^\T_\cap} 
\newcommand{\dellop}{\ominus^\T_\cap} 
\newcommand{\ins}[2]{#1 \insop #2} 
\newcommand{\insf}[1]{#1 \insop F} 
\newcommand{\dell}[2]{#1 \dellop #2} 
\newcommand{\dellf}[1]{#1 \dellop F} 
\newcommand{\insopcar}{\oplus^\T_{CAR}} 
\newcommand{\dellopcar}{\ominus^\T_{CAR}} 
\newcommand{\inscar}[2]{#1 \insopcar #2} 
\newcommand{\insfcar}[1]{#1 \insopcar F} 
\newcommand{\dellcar}[2]{#1 \dellopcar #2} 
\newcommand{\dellfcar}[1]{#1 \dellopcar F} 
\newcommand{\clos}[2]{cl_{#1}(#2)}
\newcommand{\closl}[3]{cl^{#1}_{#2}(#3)}
\newcommand{\Mod}[1]{\mathit{Mod}{(#1)}}
\newcommand{\clc}[1]{\textsf{clc}_{\T}(#1)}
\newcommand{\cl}[1]{\textsf{cl}_{\T}(#1)}
\newcommand{\clt}[1]{\textsf{cl}(#1)}
\newcommand{\TRUE}{\mathit{True}}
\newcommand{\FALSE}{\mathit{False}}
\newcommand{\unsat}{\mathit{Unsat}}

\newcommand{\person}{\textsf{Person}}
\newcommand{\man}{\textsf{Man}}
\newcommand{\woman}{\textsf{Woman}}
\newcommand{\male}{\textsf{Male}}
\newcommand{\female}{\textsf{Female}}
\newcommand{\young}{\textsf{Young}}
\newcommand{\adult}{\textsf{Adult}}
\newcommand{\kid}{\textsf{Kid}}
\newcommand{\hasfather}{\textsf{hasFather}}
\newcommand{\animal}{\textsf{Animal}}
\newcommand{\taylor}{taylor}
\newcommand{\sam}{sam}
\newcommand{\tom}{tom}
\newcommand{\bill}{bill}
\newcommand{\hasparent}{\textsf{hasParent}}

\newcommand{\id}[2]{(\mathsf{id}\; #1\; #2)}
\newcommand{\test}[1]{#1?}
\newcommand{\match}{\mathit{Match}}
\newcommand{\homeTeam}{\mathit{homeTeam}}
\newcommand{\visitorTeam}{\mathit{visitorTeam}}
\newcommand{\playedMatch}{\mathit{playedMatch}}
\newcommand{\sysubset}[2]{#1 \prec_{\Rn} #2}
\newcommand{\Rn}{Rn}
\newcommand{\suba}{\sigma_{\A}\xspace}
\newcommand{\MinIncSet}{\mathit{MinIncSet}}
\newcommand{\IncRewr}{\mathit{IncRewr}}
\newcommand{\IncRewrUCQ}{\mathit{IncRewrUCQ}}
\newcommand{\unify}{\textsf{Saturate}}
\newcommand{\compset}{\mathit{CompSet}}
\newcommand{\mvset}{minIncSets\xspace}
\newcommand{\con}{contr\xspace}
\newcommand{\satq}{\textsf{unsatQueries}}
\newcommand{\inQueries}{\textsf{InQueries}\xspace}
\newcommand{\mini}{\textsf{Min}\xspace}
\newcommand{\minsatq}{\textsf{minUnsatQueries}\xspace}
\newcommand{\satisfiableden}{\textsf{Satisfiable}_{DA}}
\newcommand{\satisfiable}{\textsf{Satisfiable}}

\newcommand{\IncSng}{\mathit{IncSng}^{\T}}
\newcommand{\ConsAtom}{\mathit{ConsAtom}}
\newcommand{\ConsAtomSet}{\mathit{ConsAtomSet}}
\newcommand{\qunsat}{\Q^{unsat}_{\T}\xspace}
\newcommand{\qsat}{\Q_{str}\xspace}
\newcommand{\qmin}{\Q^{min}_{\T}\xspace}
\newcommand{\clnt}[1]{\mathit{cln}(#1)}
\newcommand{\qsingle}{\Q^{singleton}_{\T}\xspace}

\newcommand{\digquonto}{DIG-\textsc{QuOnto}\xspace}
\newcommand{\protege}{Prot\'eg\'e\xspace}

\newcommand{\val}[1]{\mathit{val}(#1)}
\newcommand{\cert}[2]{\mathit{cert}(#1,#2)}
\newcommand{\sem}[2]{\mathit{sem}_{#1}(#2)}
\newcommand{\IDC}{\T_{\mathit{id}}}
\newcommand{\GammaVO}{\Gamma_{\mathit{VO}}}
\newcommand{\vc}{\vec{c}}
\newcommand{\chasei}[1]{\mathit{chase}_{#1}(\O)}
\newcommand{\can}[1][\K]{\mathit{can}(#1)}
\newcommand{\cani}[1]{\mathit{can}_{#1}(\O)}
\newcommand{\img}{\G}
\newcommand{\qv}[1]{\texttt{#1}}
\newcommand{\qo}[1]{\textsl{#1}}

\newcommand{\ifdirection}{(\Leftarrow)}
\newcommand{\onlyifdirection}{(\Rightarrow)}

\newcommand{\DR}{\textsf{Driver}}
\newcommand{\TM}{\textsf{TeamMember}}
\newcommand{\MC}{\textsf{Mechanic}}
\newcommand{\drives}{\textsf{drives}}
\newcommand{\FOT}{FT}

\newcommand{\notincoherent}{\mathit{ConsAtom}}
\newcommand{\disjointattrdomains}{\mathit{DisjAttrDom}}
\newcommand{\disjointconceptnames}{\mathit{DisjConcepts}}
\newcommand{\disjointroledomains}{\mathit{DisjRoleDom}}
\newcommand{\disjointroleranges}{\mathit{DisjRoleRan}}
\newcommand{\disjointrolenames}{\mathit{DisjRoles}}
\newcommand{\disjointinverseroles}{\mathit{DisjInvRoles}}
\newcommand{\disjointattributes}{\mathit{DisjAttributes}}
\newcommand{\notdisjclash}{\mathit{NotDisjClash}}
\newcommand{\notfunctclash}{\mathit{NotFunctClash}}
\newcommand{\notclash}{\mathit{NotClash}}
\newcommand{\iarincrewriting}{\textit{IncRewriting}_{\iar}}
\newcommand{\iarincrewritingUCQ}{\textit{IncRewritingUCQ}_{\iar}}
\newcommand{\icarincrewritingUCQ}{\textit{IncRewritingUCQ}_{\icar}}
\newcommand{\icarincrewriting}{\textit{IncRewriting}_{\icar}}
\newcommand{\iarrewriting}{\textit{PerfectRef}_{\iar}}
\newcommand{\icarrewriting}{\textit{PerfectRef}_{\icar}}
\newcommand{\aczero}{\mathit{AC}^0}
\newcommand{\pidue}{\Pi^p_2}

\newcommand{\emp}{\textit{employee}\xspace}
\newcommand{\mgr}{\textit{manager}\xspace}
\newcommand{\worksfor}{\textit{WORKS-FOR}\xspace}
\newcommand{\tempemp}{\textit{tempEmp}\xspace}
\newcommand{\until}{\textsf{until}\xspace}
\newcommand{\proj}{\textit{project}\xspace}
\newcommand{\manages}{\textit{MANAGES}\xspace}
\newcommand{\xsddate}{\texttt{xsd:date}\xspace}
\newcommand{\palm}{\textbf{Palm}\xspace}
\newcommand{\lenz}{\textbf{White}\xspace}
\newcommand{\projquonto}{\textbf{DIS-1212}\xspace}
\newcommand{\projtones}{\textbf{FP6-7603}\xspace}
\newcommand{\projname}{\mathsf{ProjName}\xspace}
\newcommand{\pname}{\mathsf{PersName}\xspace}
\newcommand{\xsdstring}{\texttt{xsd:string}\xspace}
\newcommand{\projectbf}{\textbf{proj}\xspace}
\newcommand{\personbf}{\textbf{pers}\xspace}
\newcommand{\managerbf}{\textbf{mgr}\xspace}
\newcommand{\term}{\textsl{25-09-05}\xspace}
\newcommand{\quontoname}{\textsl{QuOnto}\xspace}
\newcommand{\tonesname}{\textsl{Tones}\xspace}
\newcommand{\lenznid}{\textsl{X11}\xspace}
\newcommand{\lenzname}{\textsl{White}\xspace}
\newcommand{\nardnid}{\textsl{X12}\xspace}
\newcommand{\nardname}{\textsl{Black}\xspace}
\newcommand{\ssna}{\textsl{20903}\xspace}
\newcommand{\salarya}{\textsl{1000}\xspace}
\newcommand{\namea}{\textsl{Rossi}\xspace}
\newcommand{\nameb}{\textsl{White}\xspace}
\newcommand{\ssnb}{\textsl{55577}\xspace}
\newcommand{\ssnc}{\textsl{29767}\xspace}

\newcommand{\device}{\textit{Device}\xspace}
\newcommand{\port}{\textit{Port}\xspace}
\newcommand{\portin}{\textit{PortIn}\xspace}
\newcommand{\portout}{\textit{PortOut}\xspace}
\newcommand{\connectedTo}{\textit{connectedTo}\xspace}
\newcommand{\of}{\textit{of}\xspace}
\newcommand{\num}{\textit{number}\xspace}

\newcommand{\flight}{\textsf{Flight}\xspace}
\newcommand{\arrival}{\textsf{arrival}\xspace}
\newcommand{\departure}{\textsf{departure}\xspace}
\newcommand{\airport}{\textsf{Airport}\xspace}
\newcommand{\company}{\textsf{Company}\xspace}
\newcommand{\city}{\textsf{City}\xspace}
\newcommand{\locatedIn}{\textsf{locatedIn}\xspace}
\newcommand{\belongsTo}{\textsf{belongsTo}\xspace}
\newcommand{\name}{\textit{name}\xspace}
\newcommand{\xsdinteger}{\texttt{xsd:integer}\xspace}

\newcommand{\red}{red\xspace}
\newcommand{\blue}{blue\xspace}
\newcommand{\green}{green\xspace}
\newcommand{\brown}{brown\xspace}
\newcommand{\Color}{\textsf{Color}\xspace}
\newcommand{\CColor}{\textsf{CoolColor}\xspace}
\newcommand{\WColor}{\textsf{WarmColor}\xspace}

\newcommand{\son}{\textsf{Son}\xspace}
\newcommand{\husband}{\textsf{Husband}\xspace}
\newcommand{\conjset}{conj\text{-}set}

\newcommand{\Ofbc}{\O_{\mathit{fbc}}}
\newcommand{\Tfbc}{\T_{\mathit{fbc}}}
\newcommand{\Afbc}{\A_{\mathit{fbc}}}
\newcommand{\League}{\textit{League}}
\newcommand{\Nation}{\textit{Nation}}
\newcommand{\Team}{\textit{Team}}
\newcommand{\Round}{\textit{Round}}
\newcommand{\Match}{\textit{Match}}
\newcommand{\PlayedMatch}{\textit{PlayedMatch}}
\newcommand{\ScheduledMatch}{\textit{ScheduledMatch}}
\newcommand{\OF}{\textit{OF}}
\newcommand{\BELONGSTO}{\textit{BELONGS-TO}}
\newcommand{\PLAYEDIN}{\textit{PLAYED-IN}}
\newcommand{\NEXT}{\textit{NEXT}}
\newcommand{\HOST}{\textit{HOST}}
\newcommand{\HOME}{\textit{HOME}}
\newcommand{\yearE}{\mathbf{year}}
\newcommand{\code}{\mathbf{code}}
\newcommand{\dateE}{\mathbf{date}}
\newcommand{\homeGoals}{\mathbf{homeGoals}}
\newcommand{\hostGoals}{\mathbf{hostGoals}}
\newcommand{\xsdnonneg}{\texttt{xsd:nonNegativeInteger}}
\newcommand{\xsdint}{\texttt{xsd:positiveInteger}}

\newcommand{\topp}{\top_P}
\newcommand{\topa}{\top_A}
\newcommand{\botc}{\bot_C}
\newcommand{\botp}{\bot_P}
\newcommand{\bota}{\bot_A}

\newcommand{\cla}{\beta\text{-}cl(\T)}
\newcommand{\clat}{cl(\T)}
\newcommand{\pse}{(\Leftarrow)}
\newcommand{\psolose}{(\Rightarrow)}
\newcommand{\pred}{\textsf{predecessors}}

\newcommand{\Cu}{\Phi_{\T}}
\newcommand{\Cd}{\Omega_{\T}}
\newcommand{\Gstar}{\G^*_{\T}}
\newcommand{\Estar}{\E^*}
\newcommand{\alphae}{\alpha(\Estar)}
\newcommand{\computeNeg}{\textsf{computeNeg}}

\newcommand{\minitab}[2][l]{\begin{tabular}{#1}#2\end{tabular}}

\newcommand{\owl}{$\S\H\R\O\I\Q$(D)}

\newcommand{\apx}[1]{#1-\textsf{oa}}
\newcommand{\sca}[1]{#1-\textsf{sca}}
\newcommand{\diff}[1]{\textsf{diff}(#1)}
\newcommand{\es}[1]{\textsf{ES}(#1)}

\def\qedfull{\hfill{\qedboxfull}   
  \ifdim\lastskip<\medskipamount \removelastskip\penalty55\medskip\fi}
\def\qedboxfull{\vrule height 4pt width 4pt depth 0pt}

\newcommand{\apxpan}[1]{Apx_{ES}(#1)}
\newcommand{\apxcalv}[1]{Apx_{SC}(#1)}
\newcommand{\apxdis}[1]{Apx_{MAX}(#1)}

\newcommand{\MASTRO}{\textsc{Mastro}\xspace}
\newcommand{\entailed}[1]{isEntailed(#1)}
\newcommand{\subes}[1]{MaxSub_{ES}(#1)}
\newcommand{\functclash}[1]{clashes(#1)}
\newcommand{\dlliteak}{\dllitea ^{(k)}}
\newcommand{\isapx}[1]{isApx(#1)}
\newcommand{\computeApx}[1]{computeApx(#1)}
\newcommand{\rolechain}{C_{\exists R_{1}...\exists R_{n}}}

\newcommand{\subs}{\mathit{subset_k}}
\newcommand{\gapx}{\mathit{globalApx}}

\newcommand{\add}[1]{\textsf{i}(#1)}
\newcommand{\del}[1]{\textsf{d}(#1)}


\newcommand{\DATALOG}{Datalog\xspace}
\newcommand{\SQL}{\textsc{sql}\xspace}
\newcommand{\SPARQL}{\textsc{Sparql}\xspace}
\newcommand{\OWLTWO}{\textsc{owl\,2}\xspace}
\newcommand{\OWLQL}{\textsc{owl\,2\,ql}\xspace}

\newcommand{\IGA}{\mathsf{IGA}}
 \newcommand{\atoms}{\textit{Atoms}}
 \newcommand{\OC}{\textit{OptCens}}
  \newcommand{\SOC}{\textit{StCens}}
 \newcommand{\gaclosure}{\textit{GA-Closure}}
 \newcommand{\atomrewrite}{\mathit{AtomRewr}}
 \newcommand{\map}{\textit{Map}}
 \newcommand{\pol}{\P}
\newcommand{\GA}{\mathbf{GA}}
\newcommand{\cens}{\C} 
\newcommand{\theory}{\mathsf{Th}} 
 \newcommand{\clga}[1]{cl^{\T}_{GA}(#1)}

\newcommand{\TQ}{\mathit{EntQ}}
\newcommand{\TQC}{\TQ}
\newcommand{\expand}{\textit{Expand}}

\newcommand{\im}{\mathit{IM}}

\newcommand{\braverefcl}{\mathit{BraveRef}}
\newcommand{\staterefcl}{\mathit{StateRef}}

\renewcommand{\perfectref}{\mathit{PerfectRef}}
\renewcommand{\unify}{\mathit{Unify}}

\newcommand{\buy}{\mathsf{buy}}
\newcommand{\contain}{\mathsf{contain}}
\newcommand{\somedrug}{\mathsf{Abc}}
\newcommand{\antiseizure}{\mathsf{Antiseizure}}
\newcommand{\pheny}{\mathsf{phenytoin}}
\newcommand{\john}{\mathsf{john}}	
\newcommand{\alice}{\mathsf{alice}}
\newcommand{\ma}{\mathsf{m}_a}
\newcommand{\mb}{\mathsf{m}_b}

\newcommand{\dyCQE}{\ensuremath{\mathsf{dynCQE}}\xspace}


\newif\ifdraft
\drafttrue
%
\ifdraft
\marginparwidth=15mm
\newcommand{\nb}[1]{\textcolor{red}{\bf!}%
	\marginpar[\parbox{15mm}{\raggedleft\scriptsize\textcolor{red}{#1}}]%
	{\parbox{15mm}{\raggedright\scriptsize\textcolor{red}{#1}}}}
\else
\newcommand{\nb}[1]{}
\fi


\title{CQE in OWL~2~QL:\\ A ``Longest Honeymoon'' Approach \\(extended version)}
\titlerunning{CQE in OWL~2~QL: A ``Longest Honeymoon'' Approach}
\author{
	Piero Bonatti\inst{1}\orcidID{0000-0003-1436-5660}
	\and
	Gianluca Cima\inst{2}\orcidID{0000-0003-1783-5605}
	\and
	Domenico Lembo\inst{3}\orcidID{0000-0002-0628-242X}
	\and
	Lorenzo Marconi\inst{3}\orcidID{0000-0001-9633-8476}
	\and
	Riccardo Rosati\inst{3}\orcidID{0000-0002-7697-4958}
	\and
	Luigi Sauro\inst{1}\orcidID{0000-0001-6056-0868}
	\and
	Domenico Fabio Savo\inst{4}\orcidID{0000-0002-8391-8049}
}

\authorrunning{P. Bonatti et al.}
%
\institute{
    Universit{\`a} di Napoli Federico II \\ \email{\{pab,luigi.sauro\}@unina.it} \and
    University of Bordeaux, CNRS, Bordeaux INP, LaBRI \\ \email{gianluca.cima@u-bordeaux.fr} \and
    Sapienza Universit{\`a} di Roma \\ \email{\{lembo,marconi,rosati\}@diag.uniroma1.it} \and
    Universit{\`a} degli Studi di Bergamo \\ \email{domenicofabio.savo@unibg.it}}
\let\oldmaketitle\maketitle
\renewcommand{\maketitle}{\oldmaketitle\setcounter{footnote}{0}}
\sloppy
\maketitle              
	\begin{abstract}
Controlled Query Evaluation (CQE) has been recently studied in the context of Semantic Web ontologies. The goal of CQE is concealing some query answers so as to prevent external users from inferring confidential information. In general, there exist multiple, mutually incomparable ways of concealing answers, and previous CQE approaches choose in advance which answers are visible and which are not.
In this paper, instead, we study a \emph{dynamic} CQE method, namely, we propose to alter the answer to the current query based on the evaluation of previous ones. We aim at a system that, besides being able to protect confidential data, is maximally cooperative, which intuitively means that it answers affirmatively to as many queries as possible; it achieves this goal by delaying answer modifications as much as possible. 
We also show that the behavior we get cannot be intensionally simulated through a static approach, independent of query history. Interestingly, for OWL 2 QL ontologies and policy expressed through denials, query evaluation under our semantics is first-order rewritable, and thus in AC$^0$ in data complexity. This paves the way for the development of practical algorithms, which we also preliminarily discuss in the paper.

\end{abstract}
	\section{Introduction}
\label{sec:introduction}

Semantic Web technologies are increasingly used to represent and link together different sources of information coming from public organizations as well as private citizens.
This information may include sensitive knowledge, e.g. medical records or social network activities, whose disclosure may affect the privacy of individuals if not adequately protected \cite{BoSa13,DBLP:journals/jair/GrauK19}. 
Furthermore, OWL~2 ontologies allow 
one to infer implicit information from explicit data, which amplifies the risk of information leakage.

One goal of confidentiality-preserving data publishing is to prevent the disclosure of sensitive information to unauthorized users while being as cooperative as possible, that is, answering queries honestly whenever this does not harm confidentiality. 
Specifically, in controlled query evaluation (CQE) \cite{Bisk00,BiBo04} the data protection policy is declaratively specified through logical formulas and 
is enforced by altering query answers through so-called censors, which either refuse to answer some queries or lie when this is needed to protect some secrets. 
In general, there exist multiple, mutually incomparable ways of concealing answers, i.e., mutually incomparable censors. 
Different works have proposed static CQE methods, where a censor is constructed (or approximated) beforehand, establishing once and for all which queries should be answered truthfully \cite{BoSa13,CKKZ15,LeRS19,CLRS20,CLMRS21}. In several cases, such approaches are not fully cooperative, because the secure view of the data is chosen without taking the users' interests into account.

Conversely, following the work of Biskup and Bonatti \cite{BiBo04b}, in this paper we introduce a dynamic CQE (\dyCQE) method that progressively decides whether being truthful or lying, based on the specific stream of queries.
Roughly speaking, the dynamic CQE approach selects, at each step, as many censors as possible, coherently with the previous answers. By doing so, it maximizes the possibility of answering the next query honestly by choosing from the current pool of censors those that allow to answer the query truthfully (if any). 

We will prove that this method satisfies the so-called ``longest honeymoon'' property, which means that, given a sequence of queries, \dyCQE returns the longest possible sequence of honest answers before lying. 
This property can be supported with several arguments. First, without any specific model of the users' intentions, the order in which queries are posed allegedly reflects their importance.
Secondly, since we cannot foresee which nor how many queries are coming in the future, answering honestly the current query (if possible) is the most cooperative possible strategy.
We will prove also that \dyCQE is optimal in a more classical sense: the set of queries honestly answered by \dyCQE is always maximal under set containment. 

After introducing the \dyCQE framework and formally investigating its general properties (Section~\ref{sec:framework}), the paper focuses on ontologies in OWL~2~QL~\cite{W3Crec-OWL-Profiles}, a tractable profile of OWL~2 designed for data-intensive applications. 
%
For this setting, in Section~\ref{sec:fo-rewritability}, we first show that the behavior of \dyCQE cannot be simulated by static CQE through data-independent modifications of the intensional components of the framework, i.e., the ontology (TBox) 
and the formulas representing the data protection policy.
It is thus necessary to devise specific techniques to implement the dynamic approach. 
To this aim, we provide a tailored query rewriting algorithm through which we show that \dyCQE query processing in OWL~2~QL is \emph{first-order rewritable}, which implies that its data complexity is in \ACz (like the evaluation of first-order sentences, i.e., SQL, queries). 
Towards practical implementations, in Section~\ref{sec:approximations}, we present a first optimization of the query reformulation technique used to prove the first-order rewritability result, based on the information acquired by the system during the interaction with users; we also present a possible approximation of the approach, should the sequence of queries become too long for our rewriting technique.
A section on related work and one on final remarks conclude the paper.


This paper is the extended version of \cite{BCLMRSS22}.
	\section{Preliminaries}
\label{sec:preliminaries}

For the technical treatment we resort to Description Logics (DLs), which are decidable fragments of First-Order (FO) logic underpinning the OWL~2  standard.
Here, we introduce the basic notions needed in this work and refer the reader to~\cite{BCMNP03} for further details. 
The languages of our interest are built from an alphabet $\Alphabet$ that consists of unary predicates (a.k.a. \emph{atomic concepts}), binary predicates (a.k.a. \emph{atomic roles}), constants (a.k.a. \emph{individual names}), and a countably infinite supply of variables.
An atom is a formula of the form $A(t)$ or $P(t_1,t_2)$, where $A$ is an atomic concept, $P$ is an atomic role, and the terms $t$, $t_1$, $t_2$ are either variables or constants. An atom is \emph{ground} if all its terms are constants.

A DL ontology $\O = \T \cup \A$ is constituted by a TBox $\T$ and an ABox $\A$, specifying intensional and extensional knowledge, respectively. In particular, in this paper we assume that the ABox is a set of ground atoms. A \emph{model} of an ontology $\O=\T\cup\A$ is a FO interpretation that satisfies all axioms in $\T$ and $\A$. $\O$ is \emph{consistent} if it has at least one model, \emph{inconsistent} otherwise, and
\emph{entails} an FO sentence $\phi$, 
denoted $\O \models \phi$, if $\phi$ is true in every model of $\O$.
Given an ABox $\A$ and a FO sentence $\phi$, we say that \emph{$\phi$ evaluates to true in $\A$} if the evaluation of $\phi$ in the Herbrand model of $\A$ is true~\cite{Lloy87}, otherwise we say that \emph{$\phi$ evaluates to false in $\A$}.
In the paper, we often refer to the set of ground atoms entailed by $\T \cup\A$, which we denote with $\cl{\A}$. 

In this work, we focus on ontologies expressed in $\dlliter$~\cite{CDLLR07}, which is the logical counterpart of 
OWL~2~QL~\cite{W3Crec-OWL2-Profiles}. In this DL, 
a role $R$ is an atomic role $P$ or its inverse $P^-$, whereas a concept $B$ takes the form $A$, $\SOMET{P}$, or $\SOMET{P^-}$. The concepts $\SOMET{P}$ and $\SOMET{P^-}$ denote the domain and the range 
of a role $P$, respectively.
A $\dlliter$ TBox $\T$ is a set of \emph{positive inclusions} of the form $B_1\ISA B_2$ or $R_1\ISA R_2$, and \emph{negative inclusions} of the form $B_1\ISA \NOT B_2$ or $R_1\ISA \NOT R_2$.


By $\conj(\vx)$ we mean a conjunction $\alpha_1\wedge \ldots \wedge \alpha_n$ of atoms where $\vx$ indicates all the variables occurring in it. Then, a Boolean Conjunctive Query (BCQ) is an existentially quantified conjunction of atoms $\exists \vx(\conj(\vx))$ and a Boolean Union of Conjunctive Queries (BUCQ) is a disjunction $q_1\vee\ldots\vee q_n$ of BCQs. 
Sometimes we write $q\in q'$ to indicate that the BCQ $q$ is one of the BCQs of the BUCQ $q'$. Note that a ground atom can be seen as a BCQ with no variables, 
and that a BCQ is a BUCQ with only one disjunct. 

Given a BCQ $q$, $\atoms(q)$ is the set of atoms occurring in $q$. 
Given two BUCQs $q_1=q_1^1\vee\ldots\vee q_1^n$ and $q_2=q_2^1\vee\ldots\vee q_2^m$, 
we denote by $q_1\wedge q_2$ the BUCQ 
\[
\begin{array}{c}
	(q_1^1\wedge q_2^1)\vee\ldots\vee(q_1^1\wedge q_2^m)\, \vee \\
	\vdots \\
	(q_1^n\wedge q_2^1)\vee\ldots\vee(q_1^n\wedge q_2^m)\, .
\end{array}
\]

We recall that entailment of BUCQs in $\dlliter$ is FO rewritable, that is, for every $\dlliter$ TBox $\T$ and BUCQ $q$, it is possible to 
compute an FO query $q_r$, called the \emph{perfect reformulation of $q$ with respect to $\T$}, such that, for each ABox $\A$, $\T \cup \A \models q$ iff $q_r$ evaluates to true in $\A$. 
We will use the algorithm $\perfectref$ presented in~\cite{CDLLR07}, which uses only positive inclusions in $\T$ as rewriting rules to compute perfect reformulations. 
We point out that the reformulation returned by $\perfectref$ is a BUCQ. The following proposition is from~\cite{CDLLR07}. 

\begin{proposition}
\label{pro:dllite-qa}
Let $\T \cup \A$ be a consistent $\dlliter$ ontology and let $q$ be a BUCQ. Then, $\T \cup \A \models q$ iff $\perfectref(q,\T)$ evaluates to true in $\A$.
\end{proposition}

Furthermore, a \emph{policy} $\P$ is a (finite) set of \emph{denials}, that is, sentences of the form $q\imp \bot$, where $q$ is a BCQ. An interpretation satisfies a denial $q\imp \bot$ iff it does not satisfy the BCQ $q$. 
%
%
We denote by $q(\P)$ the BUCQ $\bigvee_{q\rightarrow\bot\in\P} q$. 

The following proposition follows from the definition of satisfaction of a denial 
and from Proposition~\ref{pro:dllite-qa}.

\begin{proposition}
\label{pro:dllite-policy-consistent}
Let $\T \cup \A$ be a consistent $\dlliter$ ontology and let $\P$ be a policy. Then, $\T \cup \P \cup \A$ is a consistent FO theory iff $\perfectref(q(\P),\T)$ evaluates to false in $\A$.
\end{proposition}


Our complexity results refer to data complexity, i.e., the complexity computed with respect to the size of the ABox only.
    \section{Framework}
\label{sec:framework}
We now introduce our framework. All definitions and properties given in this section apply to any DL language.

A \emph{CQE specification} is a pair $\tup{\T,\P}$, where $\T$ is a TBox and $\P$ is a policy, such that $\T \cup \P$ is consistent.
A CQE instance is a triple $\E=\tup{\T,\P,\A}$, where $\tup{\T,\P}$ is a CQE specification, and $\A$ is an ABox such that $\T \cup \A$ is consistent.

Censors specify which consequences of an ontology can be disclosed without violating the policy. The following definition is adapted from~\cite[Definition 1]{CLMRS21}.\footnote{Other definitions of censors have been considered in the literature, for example in~\cite{CKKZ15,LeRS19}. Definition~\ref{def:censor} is chosen because it yields several important properties, such as \emph{indistinguishability} (cf.\ Section~\ref{sec-related-work}), and it has been thoroughly investigated in various settings (e.g., in~\cite{CLMRS20,CLMRS21}).}

\begin{definition}[Censor]\label{def:censor}
    Let $\E=\tup{\T,\A,\pol}$ be a CQE instance. A \emph{censor} for $\E$ is an ABox $\cens \subseteq \cl{\A}$
	such that $\T \cup \P \cup \cens$ is consistent.
\end{definition}

Given a CQE instance $\E$ and a censor $\cens$ for $\E$, we say that $\cens$ is \emph{optimal} if there exists no censor $\cens'$ for $\E$ such that $\cens \subset \cens'$.
We denote by $\OC(\E)$ the set of all the optimal censors for $\E$. We observe that a censor for a CQE instance $\E$ always exists,\footnote{Trivially, the empty set is a censor for any CQE instance $\E$.} and thus $\OC(\E) \neq \emptyset$.
Given a BUCQ $q$, we denote by $\OC(\E,q)$ the set of optimal censors that, together with $\T$, entail $q$:
$$
\OC(\E,q) = \{ \cens \in\OC(\E) \mid \T \cup \cens \models q \}
$$

The following notion of \textit{protection state} captures the history of queries submitted by the users to a CQE instance.

\begin{definition}[State]
Let $\E=\tup{\T,\pol,\A}$ be a CQE instance. 
A \emph{protection state of $\E$} (or simply state of $\E$) is a pair $\S = \tup{\E, \Q}$, where $\Q=\tup{q_1,\ldots,q_n}$ (with $n\geq 0$) is a sequence of BUCQs.
\end{definition}


Below we formalize our idea of dynamic CQE (\dyCQE), i.e., a CQE that takes into account the sequence of queries that have been already processed. In what follows, given a CQE instance $\E$, a sequence $\Q_n=\tup{q_1,\ldots,q_n}$ of BUCQs, and any integer $i\in[0,n]$, we denote with $\Q_i$ the sequence $\tup{q_1,\ldots,q_i}$ and with $\S_i$ the state $\tup{\E,\Q_i}$ of $\E$, with the convention that $\Q_0$ is the empty sequence $\tup{}$.

\begin{definition}[Dynamic CQE -- \dyCQE]\label{def:censor-of-state}
Let $\E=\tup{\T,\pol,\A}$ be a CQE instance, and let $\Q_n=\tup{q_1,\ldots,q_n}$ (with $n \geq 0$) a sequence of BUCQs. The set $\SOC(\S_n)$ of censors of $\S_n$ is inductively defined as follows:
\begin{itemize}
\item $\SOC(\S_0)=\OC(\E)$; 
\item $\SOC(\S_{i+1})=\begin{cases} 
            \SOC(\S_i) \text{~if~} \SOC(\S_i)\cap\OC(\E,q_{i+1})=\emptyset,\\
            \SOC(\S_i)\cap\OC(\E,q_{i+1}) \text{~otherwise},\\
    \end{cases}$\\
    for every $0 \leq i \leq n-1$.
\end{itemize}
For each BUCQ $q_i$ occurring in $\Q_n$, we say that \emph{$q_i$ is entailed by $\S_n$}, denoted by $\S_n \models q_i$, if $\T\cup\cens\models q_i$ for every $\cens \in \SOC(\S_n)$. We denote by $\TQ(\S_n)$
the set of queries of $\Q_n$ entailed by $\S_n$, i.e., $\TQ(\S_n)=\{q\in\Q_n \mid \S_n\models q \}$.
\end{definition}

One can see that, for any $i=1,\ldots,n$, the set of censors of a state $\S_i$ is always non-empty and consists of
a subset of the set of censors in its predecessor state $\S_{i-1}$, i.e. $\SOC(\S_{i-1}) \supseteq \SOC(\S_i) \supset \emptyset$. This also means that $\TQ(\S_{i-1}) \subseteq \TQ(\S_{i})$ holds for any $i=1,\ldots,n$.

Informally speaking, each set $\SOC(\S_i)$ (with $1\le i\le n$) in the above definition progressively selects the set of optimal censors of $\E$ that agree with $\TQ(\S_i)$. 
Clearly, if 
none of the surviving optimal censors in $\SOC(\S_i)$ entails (together with $\T$) a query $q_{i+1}$, then $q_{i+1} \not \in \TQ(\S_{i+1})$, so we have that $\SOC(\S_{i+1})=\SOC(\S_i)$.
On the other hand, if at least one of the surviving censors in $\SOC(\S_i)$, together with the TBox, entails $q_{i+1}$, then, according to \dyCQE, we have a positive answer, and $\SOC(\S_{i+1})$ keeps only the censors in $\SOC(\S_i)$ that agree with this answer.

As a result, the stream of queries is processed greedily, answering the truth as long as some of the censors in $\SOC(\S_n)$ allows to do it (\textit{longest honeymoon approach}~\cite{BiBo04b}), as we will formally show below.

Note that, by Definition~\ref{def:censor-of-state}, given a state $\S=\tup{\E,\Q}$ and
a query $q$ occurring in $\Q$, we have that either $\T \cup \cens \models q$ for every $\cens \in \SOC(\S)$, or $\T\cup\cens \not \models q$ for every $\cens \in \SOC(\S)$. This means that $\S \models q$ if and only if there exists a censor $\cens \in\SOC(\S)$ such that $\T \cup \cens \models q$.

\begin{example}\label{ex:ex1}
	Some pharmaceutical products may reveal with high accuracy which kind of disease is affecting a person. For instance, drugs that contain phenytoin, or that are classified as anti-seizure medications, indicate some form of epilepsy.

	Let $\E=\tup{\T,\P,\A}$ be a CQE instance, where:
	\begin{tabbing}
	 $\T=\{\somedrug \ISA \antiseizure\}$;\\
	 $\P=\{$\=$\exists x,y (\buy(x,y)\wedge \antiseizure(y))\imp \bot,$\\ 
	        \>$\exists x,y (\buy(x,y)\wedge \contain(y,\pheny))\imp \bot\}$;\\ 
	 $\A=\{\buy(\john,\ma), \somedrug(\ma), \buy(\alice,\mb), \contain(\mb,\pheny)\}$. 	
	\end{tabbing}
	In words, the TBox states that $\somedrug$ is an anti-seizure medication, while the policy conceals the presence of patients suffering from epilepsy.

    Let us start by considering an empty sequence of BUCQs. 
    By definition, we have that $\SOC(\tup{\E,\tup{}})$ coincides with the set of the optimal censors for $\E$:
    \begin{compactitem}
    	\item $\cens_1=\{\buy(\john,\ma), \buy(\alice,\mb)\}$;
    	\item $\cens_2=\{\buy(\john,\ma), \contain(\mb,\pheny)\}$;
    	\item $\cens_3=\{\somedrug(\ma), \antiseizure(\ma), \buy(\alice,\mb)\}$;
    	\item $\cens_4=\{\somedrug(\ma), \antiseizure(\ma), \contain(\mb,\pheny)\}$.
    \end{compactitem}
    
    Let $q_1 = \buy(\john,\ma)$ be the first query. 
    The censors $\C_1$ and $\C_2$ agree with answering $\true$ to this query. All the censors that disagree with such answer are then removed, obtaining $\SOC(\tup{\E,\tup{q_1}})=\SOC(\tup{\E,\tup{}})\cap \OC(\E,q_1)=\{\cens_1, \cens_2\}$.
    %
    %
    Then, let $q_2 = \somedrug(\ma)$ be a new query in the sequence. Since neither $\T \cup \cens_1$ nor $\T \cup \cens_2$ entail $q_2$, then $\SOC(\tup{\E,\tup{q_1,q_2}})=\SOC(\tup{\E,\tup{q_1}})$. 
    Now, consider to add $q_3 = \exists x\buy(x,\mb)$ to the sequence. Since $\T \cup \cens_1 \models q_3$ while $\T \cup \cens_2 \not \models q_3$, we have $\SOC(\S) = \{\cens_1\}$, where $\S=\tup{\E,\Q}$ with $\Q=\tup{q_1,q_2,q_3}$. 
    %
    %
    %
    Clearly, $\S \models q_1$ and $\S \models q_3$, but $\S \not\models q_2$. \qed 
\end{example}	



Let $\E = \tup{\T, \P,\A}$ be a CQE instance. For all states $\S$ of $\E$, our dynamic CQE method is \emph{optimal with respect to $\S$}, in the sense that we have that $\TQ(\S)$ is never strictly contained in the set of queries of $\S$ entailed by any censor $\cens$ for $\E$. In order to formalize this property, for all states $\S=\tup{\E,\Q}$ and all censors $\cens$ for $\E$, let $\TQC(\Q,\cens,\T)$ be the subset of queries of $\Q$ entailed by $\cens\cup\T$, i.e. $\TQC(\Q,\cens,\T) = \{q\in\Q\mid \T\cup\cens\models q\}$. 

\begin{proposition}
\label{TQ-optimality}
Let $\E = \tup{\T,\P,\A}$ be a CQE instance, $\Q=\tup{q_1,\ldots,q_n}$ (with $n\geq 0$) be a sequence of BUCQs, and $\S=\tup{\E,\Q}$.
There exists no censor $\cens\in\OC(\E)$ such that $\TQ(\S) \subset \TQC(\Q,\C,\T)$.
\end{proposition}
\begin{proof}
    By contradiction, let such a censor $\C$ exist and let $i$ be the least index such that $\T\cup\cens\models q_i$ and $q_i\not\in\TQ(\S)$. By the minimality of $i$ we have that, for all $j \in \{1,\ldots,i-1\}$, $\T\cup\cens\models q_j$ iff $q_j \in \TQ(\tup{\E,\tup{q_1,\ldots,q_{i-1}}})$. It follows that $\cens\in\SOC(\tup{\E,\tup{q_1,\ldots,q_{i-1}}})$.
    But then, by definition, we should have that $\C \in \SOC(\tup{\E,\tup{q_1,\ldots,q_i}})$, and, consequently, that $q_i \in \TQ(\tup{\E,\tup{q_1,\ldots,q_i}}) \subseteq \TQ(\S)$ (a contradiction).\qed
\end{proof}

\noindent
Moreover, \dyCQE 
is the \emph{only} way to guarantee that such optimality is preserved in the future. One might object that answering the current query $q$ honestly may prevent the system from answering honestly another set of queries $\Q'$ in the future. However, the queries in $\Q'$ might never be submitted, so any censor that conceals the answer to $q$ now might remain sub-optimal in the future. This may happen no matter how many additional queries are submitted by the users.
Formally, we have:

\begin{proposition}
\label{thm-riccardo's-prop}
Let $\E=\tup{\T,\P,\A}$ be a CQE instance, $\Q=\tup{q_1,\ldots,q_n}$ 
be a sequence of BUCQs, and $\S=\tup{\E,\Q}$.
For all BUCQs $q_{n+1}$, and for all censors $\cens$ in $\SOC(\S)\setminus\SOC(\tup{\E,\Q\circ\tup{q_{n+1}}})$\footnote{With $\Q\circ\tup{q_{n+1}}$ we denote the sequence $\tup{q_1,\ldots,q_n,q_{n+1}}$.}, there exist queries $q_{n+2},q_{n+3},\ldots,q_{n+k},\ldots$ such that $\TQC(\tup{q_1,\ldots,q_i},\cens,\T) \subset\TQ(\tup{\E,\tup{q_1,\ldots,q_i}})$ for all $i>n$.
\end{proposition}
\begin{proof}
   (Sketch)
   It follows from the hypothesis that $\cens\not\in \OC(\E,q_{n+1})$, therefore $\TQC(\tup{q_1,\ldots,q_{n+1}},\cens,\T) = \TQ(\tup{\E,\tup{q_1,\ldots,q_{n+1}}})\setminus \{q_{n+1}\} \subset \TQ(\tup{\E,\tup{q_1,\ldots,q_{n+1}}})$. The simplest case in which this strict inclusion is preserved for all future queries is where the queries $q_{n+1+k}$ are picked from $\{q_1,\ldots,q_{n+1}\}$ (or logically equivalent formulae), for all $k>0$. In this way, for all $k>0$, $\TQC(\tup{q_1,\ldots,q_{n+1+k}},\cens,\T) =\TQC(\tup{q_1,\ldots,q_{n+1}},\cens,\T)$ and $\TQ(\tup{\E,\tup{q_1,\ldots,q_{n+1+k}}}) =\TQ(\tup{\E,\tup{q_1,\ldots,q_{n+1}}})$. \qed
\end{proof}


\noindent
In the above proposition, the hypothesis $\cens\in\SOC(\S)\setminus\SOC(\tup{\E,\Q\circ\tup{q_{n+1}}})$ implies that $q_{n+1}$ can be given a positive answer without disclosing any protected data, but $\cens$ does not allow a positive answer to $q_{n+1}$.


Another property of dynamic CQE is that the first answer modification occurs as late as possible (\emph{longest honeymoon} property). The following notion of maximal cooperativity implies and strengthens the longest honeymoon property.

\begin{definition}[Cooperativity]\label{def:cooperativity}
    Let $\E = \tup{\T, \P, \A}$ be a CQE instance, $\Q=\tup{q_1,\ldots,q_n}$ (with $n\geq 0$) a sequence of BUCQs, and $\cens$ and $\cens'$ two censors for $\E$. We say that $\cens$ is \emph{more cooperative than} $\cens'$ with respect to $\Q$ if there exists a non-negative natural number $m < n$ such that
    \begin{compactitem}
        \item $\T \cup \cens \models q_i \Longleftrightarrow \T \cup \cens' \models q_i$ for every $1 \leq i \leq m$, and
        \item $\T \cup \cens \models q_{m+1}$ and $\T \cup \cens' \not \models q_{m+1}$.
    \end{compactitem}
    
    We also say that $\cens$ is \emph{maximally cooperative} with respect to $\Q$ if there does not exist any censor $\cens''$ for $\E$ that is more cooperative than $\cens$.
\end{definition}
%

%

The following intermediate result shows that a state of a CQE instance cannot discriminate between two optimal censors 
if they have answered all the queries posed so far in the same way.
\begin{lemma}\label{lem:preliminary}
Let $\E = \tup{\T, \P ,\A}$ be a CQE instance, $\Q=\tup{q_1,\ldots,q_n}$ (with $n\geq 0$) be a sequence of BUCQs, and $\cens$ and $\cens'$ be two optimal censors for $\E$ such that $\T \cup \cens \models q_i \Longleftrightarrow \T \cup \cens' \models q_i$, for all $i\in\{1,\ldots, n\}$. Then, $\cens\in \SOC(\tup{\E, \Q})$ iff $\cens'\in \SOC(\tup{\E, \Q})$.
\end{lemma}
\begin{proof}
    The proof is by induction on the length of $\Q$.
	
	Case $n=0$. Since $\Q$ is empty, both $\cens$ and $\cens'$ are 
	in $\SOC(\tup{\E, \Q})$.
	
	Case $n\ge 1$. In this case $\Q=\Q'\circ \tup{q_n}$, where $\Q'=\tup{q_1,\ldots,q_{n-1}}$.
	From the assumption $\T \cup \cens \models q_i $ iff $ \T \cup \cens' \models q_i$, for all $i\in\{1,\ldots, n\}$, the following two facts hold: $(i)$ $\cens\in \OC(\E,q_n)$ iff $\cens'\in \OC(\E,q_n)$; $(ii)$ by IH, $\cens\in \SOC(\tup{\E, \Q'})$ iff $\cens'\in \SOC(\tup{\E, \Q'})$. Then, since $\SOC(\tup{\E,\Q})$ is by Definition~\ref{def:censor-of-state} equal either to $\SOC(\tup{\E,\Q'})$ or to $\SOC(\tup{\E,\Q'})\cap\OC(\E,q_n)$, we have the thesis.
	\qed
\end{proof}

Then, we prove that for all states $\S=\tup{\E,\Q}$ of a CQE instance, the set $\SOC(\S)$ coincides with the set of all censors that are maximally cooperative with respect to $\Q$.

\begin{theorem}\label{thm:we-are-maximally-cooperative}
Let $\E = \tup{\T,\P,\A}$ be a CQE instance, and $\Q=\tup{q_1,\ldots,q_n}$ (with $n\geq 0$) be a sequence of BUCQs. A censor $\cens$ for $\E$ is maximally cooperative with respect to $\Q$ iff $\cens \in \SOC(\tup{\E, \Q})$.
\end{theorem}
\begin{proof}
	We start by showing that every $\cens \in \SOC(\tup{\E,\Q})$ is maximally cooperative with respect to $\Q$. 
	Let $\S_h=\tup{\E,\tup{q_1,\ldots,q_h}}$, with $h\le n$, and assume by contradiction that, for some $\cens\in \SOC(\S_n)$, there exists an optimal censor $\cens'$ and a number $m< n$ such that $(i)$ $\T \cup \cens \models q_i \Longleftrightarrow \T \cup \cens' \models q_i$, for each $i \leq m$, and $(ii)$ $\T \cup \cens \not \models q_{m+1}$ and $\T \cup \cens' \models q_{m+1}$. 
	
	Note that the sets $\SOC(\S_h)$ form by construction a descending $\subseteq$-chain, hence $\cens$ is in $\SOC(\S_m)$. 
	Then, from $(i)$ and Lemma~\ref{lem:preliminary}, $\cens' \in \SOC(\S_m)$ too.

	From $(ii)$ we have that $\cens'$ occurs in $\OC(\E,q_{m+1})$ whereas $\cens$ does not. Then, on the one hand, since $\cens'\in \SOC(\S_m)\cap \OC(\E,q_{m+1})$, $\SOC(\S_{m+1})$ is equal by definition to $\SOC(\S_m)\cap \OC(\E,q_{m+1})$. On the other hand, $\SOC(\S_{m+1})$ does not contain $\cens$, as $\cens$ is not in $\OC(\E,q_{m+1})$. But this means that also $\SOC(\S_n)$ does not contain $\cens$, a contradiction.
	
	Now, we show that if a censor $\cens$ for $\E$ is maximally cooperative w.r.t. $\Q$, then $\cens \in \SOC(\tup{\E,\Q})$. 
	By contradiction, assume that $\cens \not \in \SOC(\tup{\E,\Q})$. So, there exists in  $\Q = \tup{q_1, \ldots, q_n}$ a query $q_i$ such that $\cens \in \SOC(\tup{\E, \tup{q_1,...,q_{i-1}}}) \setminus \SOC(\tup{\E, \tup{q_1,...,q_i}})$. Hence, there exists a censor $\cens' \in \SOC(\tup{\E, \tup{q_1,...,q_i}})$ such that $\T \cup \cens' \models q_{i}$, while $\T \cup \cens \not \models q_i$ and such that $\T \cup \cens' \models q_{j} \Longleftrightarrow \T \cup \cens \models q_{j}$ for every $1 \leq j \leq i-1$. So, by Definition~\ref{def:cooperativity}, $\cens'$ is more cooperative than $\cens$, which contradicts the fact that $\cens$ is maximally cooperative.
	\qed
\end{proof}

	

%

We conclude this section by comparing our new semantics of entailment with some other semantics from the literature. 
A first proposed strategy is arbitrarily choosing an optimal censor~\cite{BiBo07,CLRS20,CKKZ13}.
%
In this case, it might happen, as also stated by Proposition~\ref{thm-riccardo's-prop}, that one looses optimality with respect to the state $\S$. For instance, if one arbitrarily picks  censor $\cens_2$ in Example~\ref{ex:ex1}, then $\TQC(\Q,\cens_2,\T) \subset \TQ(\S)$. On the other hand, when the chosen censor $\cens$ turns out to be optimal with respect to a state $\S$, then, due to Theorem~\ref{thm:we-are-maximally-cooperative}, either $\cens \in \SOC(\S)$ or $\cens$ is not maximally cooperative with respect to $\Q$. 
%

Other two CQE semantics proposed in literature are: $(i)$ \emph{skeptical reasoning}~\cite{CLRS20,LeRS19}, where a query $q$ is entailed by a CQE instance $\E=\tup{\T,\P,\A}$, denoted by $\E \models q$, if it is entailed by all the optimal censors for $\E$ together with the TBox, i.e., $\T \cup \C \models q$ for each $\cens \in \OC(\E)$, and $(ii)$ its approximation, called $\IGA$ semantics~\cite{CLMRS20}, under which $q$ is entailed -- in symbols, $\E \models_\IGA q$ -- if it is entailed by $\T \cup \C_\IGA$, where $\C_\IGA$ is the 
intersection of all the optimal censors for $\E$, i.e, $\cens_\IGA = \bigcap_{\cens \in \OC(\E)} \C$.
The following proposition shows that skeptically reasoning over all optimal censors is always a sound approximation of \dyCQE. 

\begin{proposition}
\label{prop:comparison}
   Let $\E = \tup{\T,\P,\A}$ be a CQE instance, $\Q=\tup{q_1,\ldots,q_n}$ (with $n\geq 0$) be a sequence of BUCQs, and $q$ be a BCQ in $\Q$. We have that 
       $\E \models_{\IGA} q \implies \E \models q \implies \tup{\E,\Q} \models q$. The converse does not necessarily hold.
\end{proposition}
\begin{proof}
   Suppose that $\E \models_{\IGA} q$. By \cite[Proposition 1]{CLMRS20}, we already know that $\E \models q$. Now, since $\E \models q$ by definition means that $\T \cup \cens \models q$ holds for each $\cens \in \SOC(\S) \subseteq \OC(\E)$, we trivially have that $\tup{\E,\Q} \models q$.
   
   As for the converse, consider Example~\ref{ex:ex1}. We have that $\tup{\E,\Q} \models q_1$ but $\E \not\models q_1$ (and thus, also $\E \not\models_\IGA q_1$) because $\T \cup \C_3 \not\models q_1$. \qed
\end{proof}

	\section{First-order Rewritability of Query Entailment}
\label{sec:fo-rewritability}



We now move to the study of computational complexity of query entailment. In this investigation,
we focus on $\dlliter$ CQE specifications, i.e., whose TBox and ABox are expressed in $\dlliter$.

A first way to solve query entailment in a state might consist in finding a reduction to the stateless CQE approach, for which algorithms 
are already known. It turns out, however, that the behavior of \dyCQE cannot be intensionally simulated by a stateless CQE instance, independent of query history.

\begin{theorem}
\label{thm-inexpressible-tbox-policy}
There exist a $\dlliter$ CQE specification $\tup{\T,\pol}$ and a BUCQ $q$ such that there exist no $\dlliter$ CQE specification $\tup{\T',\pol'}$ such that, for every ABox $\A$, $\OC(\tup{\T',\P',\A})=\SOC(\S)$, where $\S=\tup{\tup{\T,\P,\A},\tup{q}}$.
\end{theorem}
\begin{proof}
Let $\T=\emptyset$, let $\P=\{C(x)\wedge D(x) \rightarrow\bot\}$, and let $q=\exists x C(x)$. By contradiction, suppose there exist a TBox $\T'$ and a policy $\P'$ such that, for every ABox $\A$, $\OC(\tup{\T',\P',\A})=\SOC(\S)$. 

Now consider the ABox $\A=\{C(a_1),C(a_2),D(a_1),D(a_2)\}$, where $a_1,a_2$ are individual names that do not appear in $\P'$. The optimal censors for $\tup{\T,\P,\A}$ are $\cens_1=\{C(a_1),C(a_2)\}$, $\cens_2=\{C(a_1),D(a_2)\}$, $\cens_3=\{D(a_1),C(a_2)\}$, $\cens_4=\{D(a_1),D(a_2)\}$. Among such optimal censors, only $\cens_4$ does not satisfy $q$. Therefore, $\SOC(\S)=\{\cens_1,\cens_2,\cens_3\}$. Since by hypothesis $\SOC(\S)=\OC(\tup{\T',\P',\A})$, it follows that 
$\T'\cup\P'\cup\cens_4$ is inconsistent and $\T'\cup\P'\cup\cens_3$ is consistent. 
Consequently, by Proposition~\ref{pro:dllite-policy-consistent}, $\perfectref(q(\P'),\T')$ evaluates to true in $\cens_4$ and 
evaluates to false in $\cens_3$.

On the other hand, it is immediate to see that, for every BUCQ $q$ that does not mention individual names in $\A$, 
$q$ evaluates to true in $\cens_4$ only if $q$ evaluates to true in $\cens_3$. Consequently, $\perfectref(q(\P'),\T')$ evaluates to true in $\cens_4$ only if $\perfectref(q(\P'),\T')$ evaluates to true in $\cens_3$. Thus we get a contradiction. \qed
\end{proof}

We now study the data complexity of the query entailment problem in a state, 
i.e., given a state $\S = \tup{\E,\Q}$ of a CQE instance $\E = \tup{\T,\P,\A}$, the problem of checking whether a BUCQ $q$ in $\Q$ belongs to $\TQ(\S)$.
In particular, we prove that this problem is FO rewritable, and, so, that it is in $\aczero$ in data complexity. 


We start by showing a fundamental property of query entailment in a state, which holds for all DLs.

\begin{theorem}
\label{thm-and-queries}
Let $\E=\tup{\T,\P,\A}$ be a CQE instance, $\Q=\tup{q_1,\ldots,q_n}$ be a sequence of BUCQs, and let $\S=\tup{\E,\Q}$. 
For every $i$ such that $1\leq i\leq n$, $q_i\in\TQ(\S)$ iff there exists a censor $\cens$ for $\E$ such that 
\[ \T \cup \cens \models \big(\bigwedge_{q\in\TQ(\S_{i-1})} q\big)\wedge q_i \]
\end{theorem}
\begin{proof}
($\Leftarrow$:) Suppose there exists a censor $\cens$ for $\E$ such that $\T \cup \cens \models (\bigwedge_{q\in\TQ(\S_{i-1})} q)\wedge q_i$.
Then, it follows immediately that there exists an optimal censor $\cens'$ for $\E$ such that $\cens'\supset\cens$, consequently $\T \cup \cens' \models (\bigwedge_{q\in\TQ(\S_{i-1})} q)\wedge q_i$. Hence, by Definition~\ref{def:censor-of-state}, $\cens' \in\SOC(\tup{\E,\tup{q_1,\ldots,q_i}})$. Therefore, $q_i\in\TQ(\S)$.

($\Rightarrow$:) Suppose $q_i\in\TQ(\S)$. Now, let $\cens'$ be an optimal censor for $\E$ such that $\cens'\in\SOC(\S)$. We have that $\T\cup\cens'\models q$ for every $q\in\TQ(\S)$, and since $q_i\in\TQ(\S)$ and $\TQ(\S_{i-1})\subseteq\TQ(\S)$, it follows that $\T \cup \cens' \models (\bigwedge_{q\in\TQ(\S_{i-1})} q)\wedge q_i$, thus proving the thesis.
\qed
\end{proof}


Given a BUCQ $q$ and an ABox $\A$, we say that an \emph{image of $q$ in $\A$} is a minimal subset $\A'$ of $\A$ such that $\A'\models q$.
Furthermore, given a BUCQ $q$, a TBox $\T$ and an ABox $\A$, we say that an \emph{image of $q$ in $\A$ with respect to $\T$} is a minimal subset $\A'$ of $\A$ such that $ \T \cup \A' \models q$.

\begin{theorem}
\label{thm:esistenza-immagine-in-chiusura}
Let $\E=\tup{\T,\P,\A}$ be a $\dlliter$ CQE instance and $\Q=\tup{q_1,\ldots,q_n}$ (with $n\geq 0$) be a sequence of BUCQs.
For every $i$ such that $1\leq i\leq n$, $q_i\in\TQ(\S)$ iff
there exists an image $\im$ of $\perfectref((\bigwedge_{q\in\TQ(S_{i-1})} q)\wedge q_i,\T)$ in $\cl{\A}$ such that $\perfectref(q(\P),\T)$ evaluates to false in $\im$.
\end{theorem}
\begin{proof}





First, we state the following property that follows straightforwardly from the definition of censor of a CQE instance and the definition of image of a BUCQ.

\begin{lemma}\label{lem:esistenza-immagine}
Let $\E = \tup{\T,\P,\A}$ be a CQE instance and $q$ be a BUCQ. There exists a censor $\cens$ for $\E$ such that $\T \cup \cens \models q$ iff there exists an image $\im$ of $q$ in $\cl{\A}$ with respect to $\T$ such that $\T \cup \P \cup \im$ is consistent.
\end{lemma}

Now, Lemma~\ref{lem:esistenza-immagine}, Proposition~\ref{pro:dllite-qa} and Proposition \ref{pro:dllite-policy-consistent} imply the following property.

\begin{lemma}
Let $\E = \tup{\T,\P,\A}$ be a CQE instance and $q$ be a BUCQ. There exists a censor $\cens$ for $\E$ such that $\T \cup \cens \models q$ iff there exists an image $\im$ of $\perfectref(q,\T)$ in $\cl{\A}$ such that $\perfectref(q(\P),\T)$ evaluates to false in $\im$.
\end{lemma}
Finally, the previous lemma and Theorem~\ref{thm-and-queries} immediately imply the theorem.
\qed
\end{proof}

Now observe that: $(i)$ $\cl{\A}$ can be computed in PTIME w.r.t.\ data complexity; $(ii)$ every image of a BUCQ $q$ has a size that is not larger than the length of the longest BCQ in $q$; $(iii)$ such a maximum length is a constant w.r.t.\ data complexity; $(iv)$ all the conditions in the theorem can be verified in PTIME with respect to data complexity~\cite{CDLLR07}. This implies that the entailment problem in a state can be decided in PTIME w.r.t.\ data complexity.

In the following, we provide a tighter upper bound, showing that this entailment problem is in $\mathit{AC}^0$ in data complexity. We do so by proving that the problem is FO rewritable.
That is, for every BUCQ $q$ of the state, there exists an FO query $q'$ that does not depend on the ABox and is such that $q$ is entailed in the state iff $q'$ evaluates to true in the ABox.

To this purpose, we will find an FO query that depends on the intensional part of the state, i.e., the TBox, the policy and the sequence of queries, and such that its evaluation on the ABox is true if and only if the condition expressed in Theorem~\ref{thm:esistenza-immagine-in-chiusura} holds (Theorem~\ref{thm:fo-rewritability-state}).
We will make two intermediate steps towards this result: first (Theorem~\ref{thm:braverefcl}), given a query $q$ on a $\dlliter$ CQE specification $\tup{\T,\P}$, we will find a query denoted by $\braverefcl(q,\T,\P)$ whose evaluation on $\cl{\A}$
corresponds to checking the existence of an optimal censor $\C$ for the CQE instance $\tup{\T,\P,\A}$ such that $\T \cup \C \models q$; 
then (Theorem~\ref{thm:fo-rewritability-state-cl}), we will find an FO query such that its evaluation on $\cl{\A}$
is true if and only if the condition expressed in Theorem~\ref{thm:esistenza-immagine-in-chiusura} holds.

Given two BCQs $q$ and $q'$, a \emph{mapping of $q'$ into $q$} is a function $h:\atoms(q')\rightarrow\atoms(q)$ such that there exists a most general unifier $\sigma_h$ such that, for every atom $\alpha\in\atoms(q')$, $\sigma_h(\alpha)=\sigma_h(h(\alpha))$. Such a most general unifier (variable substitution) assigns variables occurring either in $q'$ or in $q$ to either variables of $q$ or constants.
We denote by $\map(q',q)$ the set of all the mappings of $q'$ into $q$.

Furthermore, we denote by $\sigma_h[q]$ the variable substitutions of $\sigma_h$ limited to variables occurring in $q$.
For instance, if $q=\exists x,y,z R(x,y,z)$, $q'=\exists x' R(x',x',a)$ (where $a$ is a constant and all other arguments are variables), then $\sigma_h=\{x'\leftarrow x, y \leftarrow x, z \leftarrow a \}$ and $\sigma_h[q]=\{y \leftarrow x, z \leftarrow a \}$.

Given two BCQs 
$q$ and $q'$, we denote by $\unify(q,q')$ the formula:
\[
\bigvee_{h\in\map(q',q)} \big(\bigwedge_{x\leftarrow t\:\in\sigma_h[q]} x=t\big)
\]

\begin{definition}
\label{def:braverefcl}
Given a BUCQ $q$, a $\dlliter$ TBox $\T$ and a policy $\P$, we define $\braverefcl(q,\T,\P)$ as the FO sentence:
\[
\bigvee_{q_r\in\perfectref(q,\T)} \exists \vx_r \big(\conj_r(\vx_r)\wedge\neg\big(\bigwedge_{q_d\in \perfectref(q(\P),\T)}\unify(q_r,q_d)\big)\big)
\]
(where we assume $q_r=\exists \vx_r (\conj_r(\vx_r))$).
\end{definition}

We now establish the fundamental property of the above query reformulation function $\braverefcl$.

\begin{theorem}
\label{thm:braverefcl}
Let $\tup{\T, \P}$ be a $\dlliter$ CQE specification. For every ABox $\A$, there exists an optimal censor $\cens$ for $\tup{\T,\P,\A}$ such that $\T \cup \cens \models q$ iff $\braverefcl(q,\T,\P)$ evaluates to true in $\cl{\A}$.
\end{theorem}
\begin{proof}
We first prove the following property.

\begin{lemma}
\label{lem:query-image-consistent}
Let $q=\exists \vx (\conj(\vx))$ and $q'=\exists \vx' (\conj'(\vx'))$ be BCQs and let $\A$ be an ABox. There exists an image $\im$ of $q$ in $\A$ such that $\im\not\models q'$ iff $\A\models \exists \vx \big(\conj(\vx) \wedge\neg\unify(q,q')\big)$, where $\unify(q,q')$ is the formula
\[
\bigvee_{h\in\map(q',q)} \big(\bigwedge_{x\leftarrow t\:\in\sigma_h[q]} x=t\big).
\]
Moreover, let $q''$ be a BUCQ. There exists an image $\im$ of $q$ in $\A$ such that $\im\not\models q''$ iff $\A\models 
\exists \vx \big(\conj(\vx)\wedge \neg(\bigwedge_{q_d\in q''}\unify(q_d,q)\big)$.
\end{lemma}
\begin{proof}
We show the thesis for BCQs $q'$, the extension to union of conjunctive queries $q''$ is a direct consequence. Moreover, we assume w.l.o.g. that $\vx$ and $\vx'$ are disjoint. 

($\Rightarrow$:) Let $\im$ be an image of $q$ in $\A$ such that $\im\not\models q'$, then there exists an answer substitution $\sigma_\im$ of the variables in $\vx$ such that $\im=\sigma_\im(\atoms(q))$, which also implies that $\A\models \sigma_\im(\conj(\vx))$. It remains to show that $\A$ does not entail $\sigma_\im(\unify(q,q'))$. By contradiction, assume $h$ to be a mapping from $q'$ to $q$ such that $\A$ entails $\sigma_\im(\bigwedge_{x\leftarrow t\:\in\sigma_h[q]} x=t)$. This implies that $\sigma_\im$ is a specialization of $\sigma_h[q]$, i.e. $\sigma_\im\circ \sigma_h[q]=\sigma_\im$. Then, for each atom $\beta\in \atoms(q')$, we have by construction that 
\[
[\sigma_\im\circ \sigma_h](\beta)=\sigma_\im(\sigma_h(\beta))=\sigma_\im(\sigma_h(h(\beta)))=[\sigma_\im\circ \sigma_h[q]](h(\beta))=\sigma_\im(h(\beta)). 
\]
Since $h(\beta)$ is an atom in $q$ and $\im$ is equal to $\sigma_\im(\atoms(q))$, $[\sigma_\im\circ \sigma_h](\beta)$ occurs in $\im$, for each $\beta\in \atoms(q')$. But this means that $\im\models q'$ against the hypothesis.

($\Leftarrow$:) 
Assume that $\A$ entails $\exists \vx \big(\conj(\vx) \wedge\neg\unify(q,q')\big)$. This means that there exists an answer substitution $\sigma$ such that (i) $\sigma(\atoms(q))\subseteq \A$, and (ii) $\A\not\models \sigma(\unify(q,q'))$. 
Let $\im= \sigma(\atoms(q))$, from (i) we know that $\I$ is an image of $q$ in $\A$.
Then, assume by contradiction that $\im\models q'$. This means that for some answer substitution $\sigma'$, $\sigma'(\atoms(q'))\subseteq \im$.
 
Let $h$ be the mapping of $q'$ into $q$ such that $h(\beta)=\alpha$ only if $\sigma'(\beta)=\sigma(\alpha)$, we know from $\sigma'(\atoms(q'))\subseteq \im$ that this mapping exists. 
The corresponding most general unifier $\sigma_h$ over $\vx$ and $\vx'$ can be defined as follows. 

For the variables $x'$ occurring in $\vx'$, $\sigma_h(x')= t$ where 
for some $\beta\in \atoms(q')$ and $h(\beta)\in \atoms(q)$, $x'$ and $t$ occurs in the same position, respectively. 

For the variables $x$ occurring in $\vx$, we have that
$$
\sigma_h(x)=
\begin{cases}
    y & \mbox{ if for some }x' \mbox{ in } \vx',\: \sigma_h(x')=y \mbox{ and }\sigma(x)=\sigma(y)\, ;\\
    \sigma(x) & \mbox{otherwise}\, .
\end{cases}
$$

Note that, for each a variable $x$ in $\vx$, either $[\sigma\circ\sigma_h[q]](x)= \sigma(\sigma(x))$ or $[\sigma\circ\sigma_h[q]](x)= \sigma(y)$ with $\sigma(x)=\sigma(y)$. In both the cases, we have that $[\sigma\circ\sigma_h[q]](x)= \sigma(x)$. But this means that $\sigma$ is a specialization of $\sigma_h$, which conflicts with (ii).
\qed
\end{proof}
We are now able to prove the theorem.

First, from Proposition~\ref{pro:dllite-qa} it follows that there exists an optimal censor $\cens$ for $\tup{\T,\P,\A}$ such that $\tup{\T,\cens}\models q$ iff there exists an optimal censor $\cens$ for $\tup{\T,\P,\A}$ such that $\cens\models\perfectref(q,\T)$.

Now, since $\cens\subseteq\cl{\A}$, and since every subset $\A'$ of $\cl{\A}$ such that $\T\cup\P\cup\A'$ is consistent is contained in some optimal censor for $\tup{\T,\P,\A}$, it follows that there exists an optimal censor $\cens$ for $\tup{\T,\P,\A}$ such that $\cens\models\perfectref(q,\T)$ iff there exists an image $I$ of $\perfectref(q,\T)$ in $\cl{\A}$ such that $\T\cup\P\cup I$ is consistent. 
Finally, by Proposition~\ref{pro:dllite-policy-consistent}, $\T\cup\P\cup I$ is consistent iff $I\not\models\perfectref(q(\P),\T)$. Consequently, Lemma~\ref{lem:query-image-consistent} implies the thesis.
\qed
\end{proof}

Then, we use $\braverefcl$ to define the new query reformulation function $\staterefcl$ as follows.

\begin{definition}
\label{def:staterefcl}
Let $\E=\tup{\T,\P,\A}$ be a $\dlliter$ CQE instance, $\Q=\tup{q_1,\ldots,q_n}$ (with $n\geq 0$) be a sequence of BUCQs, let $i$ be such that $1\leq i\leq n$,
and let $I\subseteq\{1,\ldots,i-1\}$: $I$ represents the set of indexes of the queries that precede query $q_i$ in $\Q$ and that are guessed to be true in the state $\S = \tup{\E,\Q}$.
We define $\staterefcl(\S,i,I)$ as the FO sentence:
\[
\big(\bigwedge_{\scriptsize\begin{array}{c}1\leq j\leq i-1 \\\wedge \;j\not\in I\end{array}}\neg\braverefcl((\bigwedge_{\ell\in I\:\wedge\:\ell<j} q_\ell)\wedge q_j,\T,\P)\big) \wedge \braverefcl((\bigwedge_{\ell\in I} q_\ell)\wedge q_i,\T,\P)
\]
\end{definition}


As an example, consider the $\dlliter$ CQE instance $\E = \tup{\T,\P,\A}$ and the query sequence $\Q = \tup{q_1, q_2, q_3}$ of Example~\ref{ex:ex1}, and let us set $i = 3$ and $I = \{ 1 \}$. We have that $\staterefcl(\tup{\E,\Q},i,I)$ is the FO sentence $\neg \braverefcl(q_1 \wedge q_2, \T, \P) \wedge \braverefcl(q_1 \wedge q_3,\T,\P) = \neg (\buy(\john,\ma) \wedge \somedrug(\ma) \wedge \neg (\exists z,w (\buy(z,w)\wedge \somedrug(w) \wedge z = \john \wedge w = \ma))) \wedge \exists x (\buy(\john,\ma) \wedge \buy(x,\mb))$.

\medskip

The query reformulation function $\staterefcl$ allows for reducing query entailment in a state to evaluating an FO query, as stated by the following property. 


\begin{theorem}
\label{thm:fo-rewritability-state-cl}
Let $\E=\tup{\T,\P,\A}$ be a $\dlliter$ CQE instance, $\Q=\tup{q_1,\ldots,q_n}$ (with $n\geq 0$) be a sequence of BUCQs.
For every $i$ such that $1\leq i\leq n$, $q_i\in\TQ(\S)$ iff the following FO sentence evaluates to true in $\cl{\A}$:
\[
\bigvee_{I\in\wp(\{1,\ldots,i-1\})}\staterefcl(\S,i,I) \text{,}
\]
where $\wp(\{1,\ldots,i-1\})$ denotes the powerset of $\{1,\ldots,i-1\}$.
\end{theorem}
\begin{proof}
We first prove the following property:
\begin{lemma}
\label{lem:staterefcl}
Let $\E=\tup{\T,\P,\A}$ be a CQE instance, $\Q=\tup{q_1,\ldots,q_n}$ be a sequence of BUCQs, 
let $i$ be such that $1\leq i\leq n$, and let $I\subseteq\{1,\ldots,i-1\}$.
Then, the FO sentence $\staterefcl(\S,i,I)$ evaluates to true in $\cl{\A}$ iff $q_i\in\TQ(\S)$ and, for each $j$ such that $1\leq j< i$, $q_j\in\TQ(\S)$ iff $j\in I$.
\end{lemma}
\begin{proof}
First, suppose that $\staterefcl(\S,i,I)$ evaluates to false in $\cl{\A}$. We have two cases:
\begin{itemize}
\item[(i)] the sentence $\braverefcl((\bigwedge_{\ell\in I} q_\ell)\wedge q_i,\T,\P)$ (that is a conjunct of $\staterefcl(\S,i,I)$) evaluates to false in $\cl{\A}$. Then, Theorem~\ref{thm:braverefcl} implies that there exists no optimal censor $\cens$ for $\E$ such that $\T\cup\C\models (\bigwedge_{\ell\in I} q_\ell) \wedge q_i$, which implies that either $q_{i}\not\in\TQ(\S)$ or there exists $j\in I$ such that $q_j\not\in\TQ(\S)$, thus proving the thesis;

\item[(ii)] the sentence $\braverefcl((\bigwedge_{\ell\in I} q_\ell)\wedge q_i,\T,\P)$ (that is a conjunct of $\staterefcl(\S,i,I)$) evaluates to true in $\cl{\A}$ and there exists $j$ such that $\leq j<i$ and $j\not\in I$ and the sentence $\neg\braverefcl((\bigwedge_{\ell\in I\:\wedge\:\ell<j} q_\ell)\wedge q_j,\T,\P))$ (that is a conjunct of $\staterefcl(\S,i,I)$) evaluates to false in $\cl{\A}$. Let us assume that $j$ is the least index such that the above property holds. Then, Theorem~\ref{thm:braverefcl}) implies that there exists an optimal censor $\cens$ for $\E$ such that $\T\cup\C\models (\bigwedge_{\ell\in I\:\wedge\:\ell<j} q_\ell)\wedge q_j$, which implies that $q_j\in\TQ(\S)$, thus proving the thesis.
\end{itemize}

Then, since for each $j$ such that $1\leq j< i$, $j\in I$ iff $q_j\in\TQ(\S)$, and since $\braverefcl((\bigwedge_{j\in I} q_j) \wedge q_{i},\T,\P)$ evaluates to true in $\cl{\A}$, Theorem~\ref{thm:braverefcl} implies that there exists an optimal censor $\cens$ for $\E$ such that $\T\cup\C\models (\bigwedge_{q\in\TQ(\S_{i-1})} q) \wedge q_{i}$, which implies that $\cens\in\SOC(\S_i)$, therefore $q_{i}\in\TQ(\S_i)$, and hence $q_{i}\in\TQ(\S)$, thus proving the thesis.

\medskip

Now suppose that $\staterefcl(\S,i,I)$ evaluates to true in $\cl{\A}$.
First, we prove by induction that, for each $j$ such that $1\leq j< i$, $j\in I$ iff $q_j\in\TQ(\S)$.

Base case ($j=1$). There are two cases: 
(i) Suppose that $j\in I$. In this case, since $\staterefcl(\S,i,I)$ evaluates to true in $\cl{\A}$, also its conjunct $\braverefcl(q_1,\T,\P)$ evaluates to true in $\cl{\A}$. Therefore, Theorem~\ref{thm:braverefcl}) implies that there exists an optimal censor $\cens$ for $\E$ such that $\T\cup\C\models q_1$, which implies that $q_1\in\TQ(\S)$. 
(ii) Suppose that $j\not\in I$. In this case, since $\staterefcl(\S,i,I)$ evaluates to true in $\cl{\A}$, also its conjunct $\neg\braverefcl(q_1,\T,\P)$ evaluates to true in $\cl{\A}$, hence $\braverefcl(q_1,\T,\P)$ evaluates to false in $\cl{\A}$, thus Theorem~\ref{thm:braverefcl} implies that there exists no optimal censor $\cens$ for $\E$ such that $\T\cup\C\models q_1$, which implies that $q_1\not\in\TQ(\S)$.

Inductive case: suppose that, for each $\ell$ such that $1\leq \ell\leq j$ (with $j<i-1$), $\ell\in I$ iff $q_\ell\in\TQ(\S)$. We prove that $\ell+1\in I$ iff $q_{\ell+1}\in\TQ(\S)$.
There are two cases:
(i) Suppose that $\ell+1\in I$. In this case, since $\staterefcl(\S,\ell+1,I)$ evaluates to true in $\cl{\A}$, also its conjunct $\braverefcl((\bigwedge_{q\in\TQ(\S_\ell)} q) \wedge q_{\ell+1},\T,\P)$ evaluates to true in $\cl{\A}$. Therefore, Theorem~\ref{thm:braverefcl} implies that there exists an optimal censor $\cens$ for $\E$ such that $\T\cup\C\models (\bigwedge_{q\in\TQ(\S_\ell)} q) \wedge q_{\ell+1}$, which implies that $\cens\in\SOC(\S_{\ell+1})$, therefore $q_{\ell+1}\in\TQ(\S_{\ell+1})$, and hence $q_{\ell+1}\in\TQ(\S)$. 
(ii) Suppose that $\ell+1\not\in I$. In this case, since $\staterefcl(\S,\ell+1,I)$ evaluates to true in $\cl{\A}$, also its conjunct $\neg\braverefcl((\bigwedge_{q\in\TQ(\S_\ell)} q) \wedge q_{\ell+1},\T,\P)$ evaluates to true in $\cl{\A}$, thus $\braverefcl((\bigwedge_{q\in\TQ(\S_\ell)} q) \wedge q_{\ell+1},\T,\P)$ evaluates to false in $\cl{\A}$. Hence, Theorem~\ref{thm:braverefcl} implies that there is no optimal censor $\cens$ for $\E$ such that $\T\cup\C\models (\bigwedge_{q\in\TQ(\S_\ell)} q) \wedge q_{\ell+1}$, which implies that $q_{\ell+1}\not\in\TQ(\S)$.
\qed
\end{proof}
Then, it is immediate to verify that the theorem is a direct consequence of Lemma~\ref{lem:staterefcl}.
\qed
\end{proof}



The query reformulations defined in the last two theorems show the FO rewritability of the problems studied on $\cl{\A}$. 
We now modify such reformulations to evaluate them directly on the ABox $\A$ and thus produce "genuine" FO rewritability results.

In what follows we will make use of the algorithm $\atomrewrite$ provided in \cite{CLRS20}, that we now briefly describe. Given an FO sentence $\phi$ and a $\dlliter$ TBox $\T$, $\atomrewrite(\phi,\T)$ computes the FO sentence obtained from $\phi$ by replacing every atom $\alpha=p(\vx)$ (where $\vx$ are all the variables occurring in $\alpha$) with the disjunction of atoms corresponding to the perfect rewriting of the non-Boolean atomic query $q_\alpha=\{\vx \mid p(\vx)\}$ with respect to $\T$. 

For our purposes, we recall the key property of $\atomrewrite$ provided in \cite{CLRS20}. 
\begin{proposition}
\label{thm:atomrewrite}
For every FO sentence $\phi$, $\dlliter$ TBox $\T$, and ABox $\A$, $\phi$ evaluates to true in $\cl{\A}$ iff $\atomrewrite(\phi,\T)$ evaluates to true in $\A$.
\end{proposition}



Now, Proposition~\ref{thm:atomrewrite} and  
Theorem~\ref{thm:fo-rewritability-state-cl} immediately imply the next property.


\begin{theorem}
\label{thm:fo-rewritability-state}
Let $\E=\tup{\T,\P,\A}$ be a $\dlliter$ CQE instance, $\Q=\tup{q_1,\ldots,q_n}$ be a sequence of BUCQs.
For every $i$ such that $1\leq i\leq n$, $q_i\in\TQ(\S)$ iff the following FO sentence evaluates to true in $\A$:
\[
\atomrewrite(\bigvee_{I\in\wp(\{1,\ldots,i-1\})}\staterefcl(\S,i,I),\T) 
\]
\end{theorem}
\begin{proof}
First, the following properties are immediate consequences of Proposition~\ref{thm:atomrewrite} together with Theorem~\ref{thm:braverefcl}, Lemma~\ref{lem:staterefcl} and Theorem~\ref{thm:fo-rewritability-state-cl}:

\begin{lemma}
\label{lem:braveref}
For every ABox $\A$, there exists an optimal censor $\cens$ for $\tup{\T,\P,\A}$ such that $\T \cup \cens \models q$ iff $\atomrewrite(\braverefcl(q,\T,\P),\T)$ evaluates to true in $\A$.
\end{lemma}

\begin{lemma}
\label{lem:stateref}
Let $\E=\tup{\T,\P,\A}$ be a CQE instance, $\Q=\tup{q_1,\ldots,q_n}$ be a sequence of BUCQs,
let $i$ be such that $1\leq i\leq n$, and let $I\subseteq\{1,\ldots,i-1\}$.
Then, the FO sentence $\atomrewrite(\staterefcl(\S,i,I),\T)$ evaluates to true in $\A$ iff $q_i \in\TQ(\S)$ and, for each $j$ such that $1\leq j< i$, $q_j\in\TQ(\S)$ iff $j\in I$.
\end{lemma}
Then, the theorem follows from the previous lemmas.
\qed
\end{proof}

The previous theorem shows the FO rewritability of the problem of entailment of BUCQs in a state.

\begin{example}
Let $\E$ and $\Q = \tup{q_1, q_2, q_3}$ be as in Example~\ref{ex:ex1}. According to Theorem~\ref{thm:fo-rewritability-state}, the query $q_3 = \exists x\buy(x,\mb)$ belongs to $\TQ(\tup{\E,\Q})$ if and only if the FO sentence below evaluates to true in $\A$ ($f_{I}$ denotes the sub-formula considering the guess $I$ of the indexes of the queries that precede the query $q_3$): 
%
%
$$
\begin{array}{l|ll}
    &  \atomrewrite( \bigvee_{I \in \wp(\{1,2\})} \staterefcl(\tup{\E,\Q},i,I), \T) = & \\ 
    f_{I = \emptyset}&\neg \braverefcl(q_1,\T,\P) \wedge \neg \braverefcl(q_2,\T,\P) \wedge \braverefcl(q_3,\T,\P)
    \vee \\ 
    f_{I = \{ 1 \}}&\neg \braverefcl(q_1 \wedge q_2, \T, \P) \wedge \braverefcl(q_1 \wedge q_3,\T,\P) 
    \vee \\
    f_{I = \{ 2 \}}&\neg \braverefcl(q_1, \T, \P) \wedge \braverefcl(q_2 \wedge q_3,\T,\P) 
    \vee\\
    f_{I = \{ 1,2 \}}&\braverefcl(q_1 \wedge q_2 \wedge q_3,\T,\P) = \\
    f_{I = \emptyset} &\neg \buy(\john,\ma) \wedge \neg \somedrug(\ma) \wedge \exists x \buy(x, \mb) 
    \vee \\ 
    f_{I = \{ 1 \}}& \neg (\buy(\john,\ma) \wedge \somedrug(\ma) \wedge \neg ( \exists z,w (\buy(z,w)\wedge \somedrug(w) \wedge \\ &\qquad z = \john \wedge w = \ma))) \wedge \exists x (\buy(\john,\ma) \wedge \buy(x,\mb)) 
    \vee \\
    f_{I = \{ 2 \}}&\neg (\buy(\john,\ma) ) \wedge (\exists x (\somedrug(\ma) \wedge \buy(x, \mb) )) 
    \vee \\
    f_{I = \{ 1,2 \}}&\exists x (\buy(\john,\ma) \wedge \somedrug(\ma) \wedge \buy(x, \mb)) \wedge \\ 
    &\qquad \neg ( \exists z,w (\buy(z,w) \wedge \somedrug(w) \wedge z = \john \wedge w = \ma )) 
\end{array}
$$
which, indeed, evaluates to true in $\A$ thanks to $f_{I = \{1\}}$.\qed
\end{example}

	\section{Towards Practical Techniques and Approximations}
\label{sec:approximations}

We now provide a simplification of the query rewriting presented in Theorem~\ref{thm:fo-rewritability-state}. In particular, in a real maximally collaborative CQE system, the answers to the queries already executed (i.e., the queries belonging to the state) can obviously be stored and re-used when the next query is submitted. This allows for greatly simplifying the structure of the FO reformulation of the query defined in Theorem~\ref{thm:fo-rewritability-state}, as shown in the following.


\begin{theorem}
\label{thm:fo-rewritability-with-memory}
Let $\E=\tup{\T,\P,\A}$ be a $\dlliter$ CQE instance, $\Q=\tup{q_1,\ldots,q_n}$ be a sequence of BUCQs, let $\S = \tup{\E,\Q}$,
let $q_{n+1}$ be a BUCQ, and let $\S'=\tup{\E,\tup{q_1,\ldots,q_n,q_{n+1}}}$. Then, $q_{n+1}$ is entailed by $\S'$ iff the following FO sentence evaluates to true in $\A$:
\[
\atomrewrite(\braverefcl((\bigwedge_{q_i\in\TQ(\S)}q_i)\wedge q_{n+1},\T,\P),\T)
\]
\end{theorem}
\begin{proof}
Suppose $\S' \models q_{n+1}$, i.e. 
$\TQ(\S')=\TQ(\S)\cup\{q_{n+1}\}$. 
By Theorem~\ref{thm:fo-rewritability-state}, 
the sentence $\psi=\atomrewrite(\staterefcl(\S',n+1,I),\T)$ evaluates to true in $\A$, where $I=\{i \mid q_i\in\TQ(\S')\}$. Consequently, the sentence $\atomrewrite(\braverefcl((\bigwedge_{q_i\in\TQ(\S)}q_i)\wedge q_{n+1},\T,\P),\T)$ is equal to the last conjunct of $\psi$, 
and therefore evaluates to true in $\A$ as well.

Suppose now $\S' \not \models q_{n+1}$. 
From Theorem~\ref{thm:fo-rewritability-state}, we have that the sentence $\atomrewrite(\staterefcl(\S',n+1,I),\T)$ evaluates to false in $\A$, where $I=\{i \mid q_i\in\TQ(\S)\}$. Since $\TQ(\S)$ is the set of BUCQ from $\tup{q_1,\ldots,q_n}$ entailed by $\S$, 
all the conjuncts of $\atomrewrite(\staterefcl(\S',n+1,I),\T)$ except the last one evaluate to true in $\A$. 
This means that its last conjunct evaluates to false in $\A$. Such a conjunct is equal to the sentence $\atomrewrite(\braverefcl((\bigwedge_{q_i\in\TQ(\S)}q_i)\wedge q_{n+1},\T,\P),\T)$, which proves the thesis.
\qed
\end{proof}

\begin{example}
Let $\E$ and the queries $q_1$, $q_2$, and $q_3$ be as in Example~\ref{ex:ex1}. Consider the sequence of queries $\Q = \tup{q_1, q_2}$. From Example~\ref{ex:ex1}, we know that only  $q_1 = \buy(\john,\ma)$ belongs to $\TQ(\tup{\E, \Q})$. Hence, according to Theorem~\ref{thm:fo-rewritability-with-memory}, the query $q_3 = \exists x\buy(x,\mb)$ is entailed by the state $\tup{\E, \Q \circ \{q_3\}}$ if and only if the FO sentence $\exists x( \buy(\john,\ma) \wedge \buy(x,\mb))$ evaluates to true in $\A$. 
%
\qed
\end{example}

An issue that the query rewriting technique of Theorem \ref{thm:fo-rewritability-with-memory} 
does not solve is the scalability w.r.t. the number of submitted queries, which might become too large to make the FO query produced by the rewriting executable in practice.
On the other hand, Theorem~\ref{thm-inexpressible-tbox-policy} shows that it is not always  possible to intensionally simulate \dyCQE by using a stateless CQE specification, i.e., through an ABox-independent transformation of the intensional part of a CQE instance.

To overcome the above issue, a possible approach
is to materialize a censor $\C$ of the current state $\S$ of the CQE instance, and then evaluate the next queries over the ontology $\T \cup \C$. 
If the current state $\S$ has multiple censors, evaluating a query over $\T \cup \C$ is only an approximation of the query entailment through \dyCQE, i.e., in the corresponding state.
More precisely: as long as the materialized system processes only queries entailed by $\T \cup \C$ (i.e., it always answers ``yes''), it returns exactly the same answers provided by \dyCQE. The first time it processes a query $q$ non-entailed by $\T \cup \C$ (i.e., it answers ``no''), its behaviour might differ from the dynamic approach, where $q$ might be either entailed or not entailed (depending on how the censors of the states evolve).
After the first negative answer, the system using $\C$ might answer ``yes'' (resp. ``no'') to a subsequent query $q$ even if the state does not entail (resp.\ entail) $q$. 
Obviously, if the state $\S$ has the only censor $\C$, then $\T \cup \C$ and the \dyCQE system will have the same behaviour. 
%
%
Below we describe how to materialize a censor of a state.
\begin{enumerate}
    \item Split the FO query of Theorem~\ref{thm:fo-rewritability-with-memory}, by executing only one 
    disjunct at a time, i.e., consider one 
    $q'\in\perfectref(\TQ(\S)\cup\{q\},\T)$ at a time, and turn all the variables appearing in $\atomrewrite(q',\T)$ as free variables. 
    \item As soon as one of such queries is true in $\A$, we can construct (through the corresponding binding of the free variables of the query) an image of this query in $\A$. Let $\A'$ be such a subset of $\A$. 
    \item $\P \cup \A'$ is consistent, so there exists at least one censor $\C$ of $\S$ that contains $\A'$. One such censor can be computed by first setting $\C = \A'$, and then, as long as it possible, by iteratively adding to $\C$ ground atoms $\gamma$ from $\cl{\A} \setminus \A'$ such that $\T \cup \P \cup \C \cup \{ \gamma \}$ is consistent.
%
%
\end{enumerate}

	\section{Related Work}
\label{sec-related-work}

As shown in ~\cite{CLMRS21}, the censors introduced in Definition~\ref{def:censor} enjoy the \emph{indistinguishability property},
that is, for all CQE instances $\E=\tup{\T,\P,\A}$ and all censors $\cens$ for $\E$, there exists an ABox $\A'$ that entails no secrets, such that $\cens$ is also a censor for $\E=\tup{\T,\P,\A'}$. 
Such censors are called \emph{indistinguishability-based} (IB) because the instances with $\A$ and $\A'$ cannot be distinguished based on the answers allowed by $\cens$. IB censors are secure against attackers that know the censor's algorithm (provided that they have no knowledge about the ABox besides the query answers returned by the censor). In particular, even if the attackers could compute the ABoxes that yield $\cens$, using their knowledge about the algorithm, the ABox $\A'$ would prevent them from inferring any secret.

Benedikt et al.\ \cite{BeCK18} provide, for OBDA settings, a systematic complexity analysis of confidentiality preserving query answering based on indistinguishability. They do not address the issue of selecting a secure data disclosure among the available ones. 
IB censors in OBDA are also considered in~\cite{CLMRS20}, where a practical approach to skeptical reasoning in CQE is presented. Differently from our approach, in~\cite{CLMRS20} censors do not take into account the history of the users' queries.

In \cite{CLRS20}, IB censors are compared with so-called \emph{confidentiality preserving} (CP) censors, that in general do not enjoy the indistinguishability property. Moreover, \cite{CLRS20} introduces algorithms and complexity results for skeptical reasoning in CQE, that is, the problem of computing only the query answers that are returned by \emph{all} IB censors. By definition, the skeptical CQE method is generally less cooperative than the dynamic method introduced and analyzed in this paper (Theorem~\ref{prop:comparison}).  In \cite{CLMR*21}, policies have been extended with numerical restrictions, and it is proved that this extension preserves FO rewritability.

The first IB CQE method for Description Logics was introduced in \cite{BoSa13}. Its confidentiality model is more robust and general, as it takes into account both object-level and meta-level background knowledge of the attacker.
However, CQ answering and FO rewritability are not addressed. Moreover, the \emph{secThere are sre views} of \cite{BoSa13} are constructed from a sequence of queries that covers \emph{all} possible relevant queries, while the properties we investigate here hold for arbitrary (possibly non-exhaustive) sequences of queries submitted by the users. 

The issue of how to select an optimal censor has been tackled in \cite{CLMRS21}. The selection criterion is based on explicit preferences over predicates, that are specified together with the CQE instance.  This approach, in general, is neither maximally cooperative nor optimal w.r.t.\ a given state, because the optimal censor is selected statically, in a stateless fashion. Moreover, the given preferences are not always able to select a single optimal censor.   

Other CQE approaches based on censors, such as CP censors, in general do not enjoy the indistinguishability property \cite{CKKZ15,LeRS19}, which makes them vulnerable to attacks based on knowledge of the CQE algorithm. Moreover, they do not address dynamic query-based censor selection. See \cite{BoSa13} for a list of earlier approaches with similar features focused on publishing secure subsets of the ontology. 


Two nice abstract analyses of censors properties can be found in \cite{StWe14,DBLP:journals/corr/abs-2005-13811}.

Finally, Cuenca Grau et al.\ \cite{DBLP:journals/jair/GrauK19} introduce and investigate an anonymization framework for knowledge graphs based on substituting nodes with blanks.
	\section{Conclusions}
\label{sec:conclusions}

In this paper, 
we have presented a maximally cooperative approach to controlled query evaluation in 
OWL and Description Logic ontologies. We have shown that the approach 
is computationally not harder than the previous static and less cooperative approaches to CQE. Moreover, we have defined a new query rewriting algorithm to solve the query entailment problem in this framework.

The present work can be extended in several interesting directions.
First, while the presented results 
indicate the possibility of a query rewriting approach to dynamic CQE, more work is still needed to define a practical query answering technique and to extend it to non-Boolean UCQs.

Then, the policy language adopted in this paper (set of denials) can be extended to encompass more expressive data protection policies. One step towards this direction, although in the context of static CQE, has been presented e.g.\ in \cite{CLMR*21}: it would be interesting to see whether \dyCQE can also be extended in a similar way.
Finally, it would be interesting to study the computational properties of dynamic CQE in ontology languages different from OWL~2 QL and $\dlliter$, in particular in the other lightweight profiles of OWL~2. 



\bibliographystyle{abbrv}
\bibliography{bibliography/short-string,bibliography/krdb,bibliography/w3c,bibliography/local-bib}

\end{document}